\documentclass{emulateapj}

\newcommand{\bvec}[1]{\textbf{#1}}

\shorttitle{Scaling Relations and Overabundance of Massive High-Redshift Clusters}
\shortauthors{Jee et al.}

\begin{document}

\title{SCALING RELATIONS AND OVERABUNDANCE OF MASSIVE CLUSTERS AT $z\gtrsim1$
FROM WEAK-LENSING STUDIES WITH {\it HUBBLE SPACE TELESCOPE} \footnotemark[1]}

\footnotetext[1]{Based on observations made with the NASA/ESA {\it Hubble Space Telescope},
obtained at the Space Telescope Science Institute}

\author{M.~J.~JEE\altaffilmark{2},
K.~S.~DAWSON\altaffilmark{3},
H.~HOEKSTRA\altaffilmark{4},
S.~PERLMUTTER\altaffilmark{5},
P.~ROSATI\altaffilmark{6},
M.~BRODWIN\altaffilmark{7},
N.~SUZUKI\altaffilmark{5}
B.~KOESTER\altaffilmark{8},
M.~POSTMAN\altaffilmark{9},
L.~LUBIN\altaffilmark{2},
J.~MEYERS\altaffilmark{5},
S.~A.~STANFORD\altaffilmark{2,17},
K.~BARBARY\altaffilmark{5,10},
F.~BARRIENTOS\altaffilmark{11},
P.~EISENHARDT\altaffilmark{12},
H.~C.~FORD\altaffilmark{13},
D.~G.~GILBANK\altaffilmark{14},
M.~D.~GLADDERS\altaffilmark{8},
A.~GONZALEZ\altaffilmark{15},
D.~W.~HARRIS\altaffilmark{3},
X.~HUANG\altaffilmark{5},
C.~LIDMAN\altaffilmark{16},
E.~S.~RYKOFF\altaffilmark{5},
D.~RUBIN\altaffilmark{5,10},
A.~L.~SPADAFORA\altaffilmark{5},
}

\altaffiltext{2}{Department of Physics, University of California, Davis, One Shields Avenue, Davis, CA 95616}
\altaffiltext{3}{Department of Physics and Astronomy, University of Utah, Salt Lake City, UT 84112}
\altaffiltext{4}{Leiden Observatory, Leiden University, Leiden, The Netherlands}
\altaffiltext{5}{E.O. Lawrence Berkeley National Lab, 1 Cyclotron Rd., Berkeley CA, 94720}
\altaffiltext{6}{European Southern Observatory, Karl-Schwarzschild-Strasse 2, D-85748, Garching, Germany}
\altaffiltext{7}{Harvard-Smithsonian Center for Astrophysics, 60 Garden Street, Cambridge, MA 02138}
\altaffiltext{8}{Department of Astronomy and Astrophysics, University of Chicago, Chicago, IL 60637}
\altaffiltext{9}{Space Telescope Science Institute, 3700 San Martin Drive, Baltimore, MD 21218, USA}
\altaffiltext{10}{Department of Physics, University of California Berkeley, Berkeley, CA 94720}
\altaffiltext{11}{Department of Astronomy and Astrophysics, Universidad Catolica de Chile, Santiago, Chile}
\altaffiltext{12}{Jet Propulsion Laboratory, California Institute of Technology, Pasadena, CA, 91109}
\altaffiltext{13}{Department of Physics and Astronomy, Johns Hopkins University, 3400 North Charles Street, Baltimore, MD 21218}
\altaffiltext{14}{Department of Physics and Astronomy, University Of Waterloo, Waterloo, Ontario, Canada N2L 3G1}
\altaffiltext{15}{Department of Astronomy, University of Florida, Gainesville, FL 32611-2055}
\altaffiltext{16}{Australian Astronomical Observatory, P O Box 296, Epping NSW 1710, Austrailia}
\altaffiltext{17}{Institute of Geophysics and Planetary Physics, Lawrence Livermore National Laboratory, Livermore, CA 94550}

\begin{abstract}
We present weak gravitational lensing analysis of 22 high-redshift ($z\gtrsim1$) clusters based on {\it Hubble Space Telescope} images.
Most clusters in our sample provide significant lensing signals and are well detected in their reconstructed two-dimensional 
mass maps. Combining the current results and our previous weak-lensing studies of five other high-$z$ clusters, we 
compare gravitational lensing masses of these clusters with other observables. We revisit the question whether the presence of the most massive
clusters in our sample is in tension with the current $\Lambda$CDM structure formation paradigm.
We find that the lensing masses are tightly correlated with the gas temperatures and establish, for the first time,
the lensing mass-temperature relation at $z\gtrsim 1$.
For the power law slope of the $M-T_X$ relation ($M\propto T^{\alpha})$, we obtain $\alpha=1.54\pm0.23$. This is consistent with
the theoretical self-similar prediction $\alpha=3/2$ and with the results previously reported in the literature for much
lower redshift samples. However, our normalization is lower than the previous results by 20-30\%, indicating that 
the normalization in the $M-T_X$ relation might evolve.
After correcting for Eddington bias and updating the discovery area with a more conservative choice,
we find that  the existence of the most massive clusters in our sample still 
provides a tension with 
the current  $\Lambda$CDM model. The combined probability of finding the four most massive clusters in this sample
after the marginalization over cosmological parameters is less than 1\%.
\end{abstract}


\section{INTRODUCTION} \label{section_introduction}

Gravitationally bound systems have been the main subjects of studies for astronomers throughout history because
they are observationally identified as discrete entities in the universe. Galaxy clusters, the largest among these, are
believed to be also the last systems detached from the Hubble expansion. Studying galaxy clusters provides
unique opportunities to test our structure formation paradigm, understand gas physics, infer the properties of
dark matter, investigate cluster galaxy evolution, and constrain cosmological parameters.

The formalism for the last of these, i.e., the use of clusters for cosmology, was pioneered by Press \& Schechter (1974), who
realized that under the assumption of Gaussian primordial density fluctuation the fraction of mass in halos more massive
than the threshold $M$ is related to the fraction of the volume in which the smoothed initial density field is above
some threshold density contrast $\delta_c$. This simple, but ingenious, approach has been shown to work
well in comparison with numerical simulations. Since this early work, many authors have extended the formalism to provide
better agreement with the recent state-of-the-art numerical simulations.

The number of clusters of a given mass, or the mass function, in the local universe constrains the combined properties of the matter
density and its fluctuation $\Omega_M\sigma_8^{0.5}$ (e.g., Bahcall \& Cen 1993; Pierpaoli et al. 2001;
Reiprich \& Boehringer 2002, and references therein) whereas the evolution of the mass function breaks this degeneracy (e.g., Bahcall
\& Fan 1998). Therefore, there has been continuous effort to extend our knowledge of the cluster mass function
to higher and higher redshift regime.
During the last decade many on-going surveys have increased the number of known of $z\gtrsim1$ clusters
by many factors (e.g., Eisenhardt et al. 2008; Boehringer et al. 2006; Gladders \& Yee 2005). 
Therefore, it is now possible to study the evolution of the mass function across a large range of redshifts.
Such a study can not only provide an important check to the results so far obtained from lower redshift samples, but
also enable a critical test on the primordial non-gaussianity (e.g., Jimenez \& Verde 2009; Sartoris et al. 2010).

Without question, one of the most crucial issues is the accurate derivation of masses for high-redshift
clusters. Indirect approaches, i.e., the estimation from velocity dispersion, X-ray properties, or Sunyaev-Zeldovich observations,
require assumptions on the dynamical state of the clusters and/or the mass versus mass proxy calibration.
By contrast, gravitational lensing is a unique tool to obtain the cluster mass without relying on any dynamical assumption.		
However, because this method requires expensive, high-resolution 
observations from space, it is not practical, if not impossible, to perform a lensing analysis for complete samples.
Instead, it is much more feasible to apply the technique to a subset of the sample, and to use the results to calibrate the relation
between masses and their proxies. 

In this paper, we present weak-lensing analysis of 22 $z\gtrsim 1$ clusters based on {\it Hubble Space Telescope images (HST)}. Our detailed
study of these clusters via their weak gravitational lensing signal will provide the aforementioned, invaluable calibration
between weak-lensing masses and other observables. In addition, the most massive among these allow us to investigate 
the behavior of the high end of the mass function 
(Hoyle et al. 2011; Holz \& Perlmutter 2010). The 22 clusters were discovered in different surveys, and do not represent a complete sample, implying that
any significant excess of massive clusters (beyond the sample fluctuation) would in fact underestimate the difference between observation and theory. 
In order to complement our sample for the investigation of the mass and mass observable relation and the
implication for the cosmology, we include five other high-redshift clusters that were previously studied via lensing.

The structure of the current paper is as follows. In \textsection\ref{section_obs}, we describe our ACS data. 
The method and the weak lensing mass reconsruction for the 22 clusters are presented in \textsection\ref{section_lensing}.
\textsection\ref{section_comparison} compares the lensing measurements with other cluster properties for the combined sample of 27 clusters and presents
the mass-temperature relation at $z>1$.
We  investigate the abundance of the most massive clusters in our study relative to the theoretical expectations within the
current paradigm of the structure formation in \textsection\ref{section_abundance} before the conclusion in \textsection\ref{section_conclusion}.
Discussions on the details of our corrections to shape measurements including point spread function (PSF) and charge transfer inefficiency (CTI)
are deferred to Appendices A and B, respectively.

We assume $(\Omega_M, \Omega_{\Lambda}, h) = (0.3, 0.7, 0.7)$ for cosmology unless explicitly stated otherwise.
All the quoted uncertainties are at the 1-$\sigma$ ($\sim68$\%) level.

\section{OBSERVATIONS} \label{section_obs}

The 22 $z\gtrsim1$ clusters were observed with ACS/WFC using the F775W and F850LP filters (hereafter referred to $i_{775}$ and 
$z_{850}$, respectively) as part of the $HST$ Cluster Supernova Survey (GO-10496, PI: Perlmutter) during
the period of 2005 July-2006 November (see Dawson et al. 2009). Three clusters among these were also observed as part of Guaranteed Time
Observation 9290 and 9919 (PI: Ford) during the periods of 2002 May-June and 2004 February-March. 
Table 1 summarizes the exposure times of the 22 clusters in $i_{775}$ and $z_{850}$.

\subsection{ACS Data Reduction}

Our reduction started with the FLT images, which are the products of the standard STScI CALACS pipeline (Hack et al. 2003).
We used the ``apsis'' pipeline (Blakeslee et al. 2003) to determine shift and rotation, correct geometric
distortion, remove cosmic rays, and create final mosaic images. The determination of the shift and rotation between
different pointings is the most critical step among these for a weak-lensing analysis because any potential misalignment
induces a coherent distortion of object shapes, mimicking gravitational lensing signals. The ``apsis'' pipeline uses high S/N astronomical objects iteratively
to measure and refine the alignment through the ``match'' program developed by Richmond (2002).
We find that the average uncertainties for shift and rotation are $\sim0.02$ pixels and $\sim0.2\arcsec$, respectively, and
thus the systematics, if any, due to the alignment errors is negligible.
The Lanczos3 (windowed sinc function) kernel was used for drizzling (Fruchter \& Hook 2002) with the native $0.05\arcsec$ pixel scale of ACS/WFC.

\subsection{Object Detection and Source Selection} \label{section_selection}

A detection image was created for each cluster by weight-averaging the two passband images via apsis. Then, SExtractor (Bertin \& Arnouts 1996)
was run in dual image mode so that objects were identified from the detection image while photometry was measured on an individual filter. This provides
common isophotal areas to both filters, which enables a robust object color measurement. We 
filtered the detection image with a Gaussian kernel that approximately matches the PSF of the instrument before looking for 5 or more contiguous
pixels above 1.5 times the sky rms.
Although our criteria were experimentally determined to minimize the return of false detections by the software, inevitably manual
identification of spurious detections (e.g., saturated stars, diffraction spikes, uncleaned cosmic
rays near field boundaries, H II regions inside nearby galaxies, etc.) is always required. 
We divided the 22 clusters into several groups and a few authors were assigned to each group to carry out this manual procedure.
A possible concern is whether or not different groups might have non-negligible biases in this false object identification, 
which may propagate into cluster mass determination. To examine the possibility of bias in the manual procedure, 
several clusters were randomly selected by one author, and compared the two false object catalogs created by this author and others.
We find that about $\sim60\%$ of false objects are shared by the two catalogs, and the disagreement is mostly attributed to
very faint objects, whose fluxes are just above the sky background. Because these extremely faint objects were discarded anyway by
our magnitude and shape measurement error criteria, the resulting difference in mass estimation and two-dimensional
mass reconstruction is far below statistical errors.

We define source galaxies as objects whose $i_{775}-z_{850}$ colors are bluer than the color of the red-sequence in each cluster while
their total $z_{850}$ magnitudes ({\tt MAG\_AUTO}) are fainter than $z_{850}=24$. We also require the source galaxies to have
ellipticity measurement error less than 0.25.
The total exposure times for the 22 clusters vary substantially and thus so does the number density of source galaxies as shown in Table 1.

\subsection{Shape Measurement}

Our shape measurement method is detailed in Jee et al. 2009. Briefly, we fit a PSF-convolved elliptical Gaussian
function to source galaxies to determine their semi-major and -minor axes. Convolution with an
elliptical Gaussian is required to measure galaxies' ellipticity before photons reach
the instrument. Because an elliptical Gaussian profile is not an unbiased representation
of true galaxies, the measurement inevitably introduces a bias. On average, the bias 
is toward underestimation of ellipticity because a Gaussian profile is steeper than that of real objects; more
extended objects are subject to larger underestimates of ellipticity. In future
analyses, it may be worth experimenting with more generalized functional forms and examining
the effect based on simulated images. Nevertheless, we stress that the amount of bias induced by the 
current shape measurement is small and to first order was corrected here using the
simulation results of Jee et al. (2007b), which shows a $\sim6$\% bias for $\gamma\gtrsim 0.5$.

We model the PSF using our library constructed from stellar field observations (Jee et al. 2007b). In order to find the matching
PSF template, we use $\sim10$ high S/N stars present in each cluster field. This matching is done for each visit, and the final PSF model is obtained after the model for each visit is shifted, rotated, and stacked.
The mean residual ellipticity is less than $\sim0.01$, and this level of accuracy is sufficient for the current cluster lensing analysis. The detailed results for each cluster can be found in Appendix \ref{appendix_psf_modeling}.

Together with the PSF, CTI may be a potential source of systematics in weak lensing. For bright stars, the trailing from deferred charges is visually apparent. We quantify the effect using cosmic rays and warm pixels, and find that the effect on the ellipticity of the galaxies that we use to extract the signal is in fact much less than the statistical noise set by the finite number of galaxies. Thus, we confirm our previous
argument (Jee et al. 2009) that the ACS CTI effect is negligible for cluster lensing analysis although it is a critical source of systematics in cosmic shear studies. 
We present the details
in Appendix \ref{appendix_cti_correction}.

\section{WEAK-LENSING ANALYSIS }\label{section_lensing}

\subsection{Mass Estimation and Two-dimensional Mass Reconstruction}

In the weak-lensing regime, the characteristic length of the lensing signal variation is 
larger than the object size. Thus, the resulting shape distortion can be linearized as follows:
\begin{equation}
\textbf{A}(\bvec{x}) =  \delta_{ij} - \frac {\partial^2 \Psi (\bvec{x})} {\partial x_i \partial x_j}
= \left ( \begin{array} {c c} 1 - \kappa - \gamma _1 & -\gamma _2 \\
                      -\gamma _2 & 1- \kappa + \gamma _1
          \end{array}   \right ),  \label{eqn_lens}
\end{equation}
\noindent
where $\textbf{A}(\bvec{x})$ is the transformation matrix $\bvec{x}^\prime = \textbf{A} \bvec{x}$, which
relates a position $\bvec{x}$ in source plane to a position $\bvec{x}^\prime$ in image plane,
and $\Psi$ is the two-dimensional lensing potential.
The convergence $\kappa$ is the surface mass density in units of critical surface mass density
\begin{equation}
 \Sigma_c= \frac{c^2}{4\pi G} \frac{D(z_s)}{D(z_l) D(z_l,z_s)} \label{eqn_sigma_c}
\end{equation}
where $D(z_s)$, $D(z_l)$, and $D(z_l,z_s)$ are the angular diameter distance from the observer to the source,
from the observer to the lens, and from the lens to the source, respectively.
The convergence $\kappa$ and the shears $\gamma_{1(2)}$ are related to the lensing potential $\Psi$ via
\begin{eqnarray}
\kappa=\frac{1}{2}  (\psi_{11}+\psi_{22}) \label{eqn_kappa},~
\gamma_1=\frac{1}{2} (\psi_{11} -\psi_{22}) \label{eqn_shear1},~
\mbox{and}~
\gamma_2=\psi_{12}=\psi_{21} \label{eqn_kappa_shear},
\end{eqnarray}
\noindent
where the subscripts on $\psi_{i(j)}$ denote partial differentiation with respect to $x_{i(j)}$.

A useful quantify to estimate masses is the reduced tangential shear defined by
\begin{equation}
 g_T  = \left< -  g_1 \cos 2\phi - g_2 \sin 2\phi \right > \label{eqn_tan_shear},
\end{equation}
\noindent
where $\phi$ is the position angle of the object with respect to the cluster center, and
$g$ is the reduced shear $\gamma/(1-\kappa)$ (valid only in the weak-lensing regime).
This reduced tangential shear profile informs us of how 
the mass density of the lens changes as a function of radius. Of course, without any lensing signal, the
resulting shear profile should vanish with fluctuations consistent  with 
shot noise.

Hoekstra (2001; 2003) demonstrates that  a cosmic shear (lensing by large scale structure)
is an important limiting factor in the accuracy of cluster masses derived from the cluster tangential shears. 
Following the formalism of Hoekstra (2003), we estimate that
the cosmic shear contribution on average is $\gamma\sim0.01$ for our sample. Although this is small compared to
the shot noise set by the finite number of source galaxies, we include the effect in our final mass uncertainty.

A potential ambiguity is the choice of the cluster center when there is disagreement between centroids determined from the distribution of cluster
galaxies, lensing
mass peaks, and X-ray emission.
If we adopt a location that
maximizes the amplitude of the tangential shear or a peak in the convergence map as the cluster center,
the derived mass will be always biased high. 
Fortunately, in previous high S/N weak-lensing studies, where the number density of background galaxies is high ($>120$ per sq. arcmin),
the weak-lensing mass peaks are in good spatial agreement with cluster galaxies. However, in many cases the X-ray
peaks are conspicuously offset, indicative of merging activity or the collisional properties of the
intracluster medium. 
In this paper we adopt the centroid defined by the cluster galaxies as the center for the construction of the tangential shear
profile. For the clusters in our sample where the statistical significance of the lensing signal is high, these centroids
also agree well with the mass peaks. In several cases, where the lensing signal is weak (the cluster is not massive
and/or the image is not sufficiently deep), we occasionally observe rather large ($>20\arcsec$) offsets between mass and galaxies.
The choice of centroid will lead to an underestimate of the cluster mass if the offsets reflect real features in the system.
We do not use the inner most data points in our mass estimation because the tangential shears at very small radii
are highly sensitive to the centroid choice. Furthermore, both observationally and theoretically, it is not clear 
how the cluster mass profile behaves near the center.

There are two popular ways in deriving the lensing mass from tangential shear: aperture-mass densitometry (Fahlman et al. 1994)
and parametrized model fitting. Aperture-mass densitometry is useful when one attempts to estimate a total projected mass
within some aperture radius without requiring an assumption on the behavior of the cluster mass profile. However, 
this approach is not practical for the current data set, which in most cases provides only $\sim3\arcmin \times 3\arcmin$
areas smaller than the virial radii of the clusters. Therefore, we use the second method to derive the cluster masses.
This method determines the parameters of analytic halo models by comparing the observed tangential shears with the expected values.
Of course, a scatter is introduced because the real cluster profile is different from the model. Nevertheless, many studies show that
this parametric approach gives consistent results with the values obtained by the former (e.g., Jee et al. 2005a; Hoekstra 2007).

\begin{figure}
\plotone{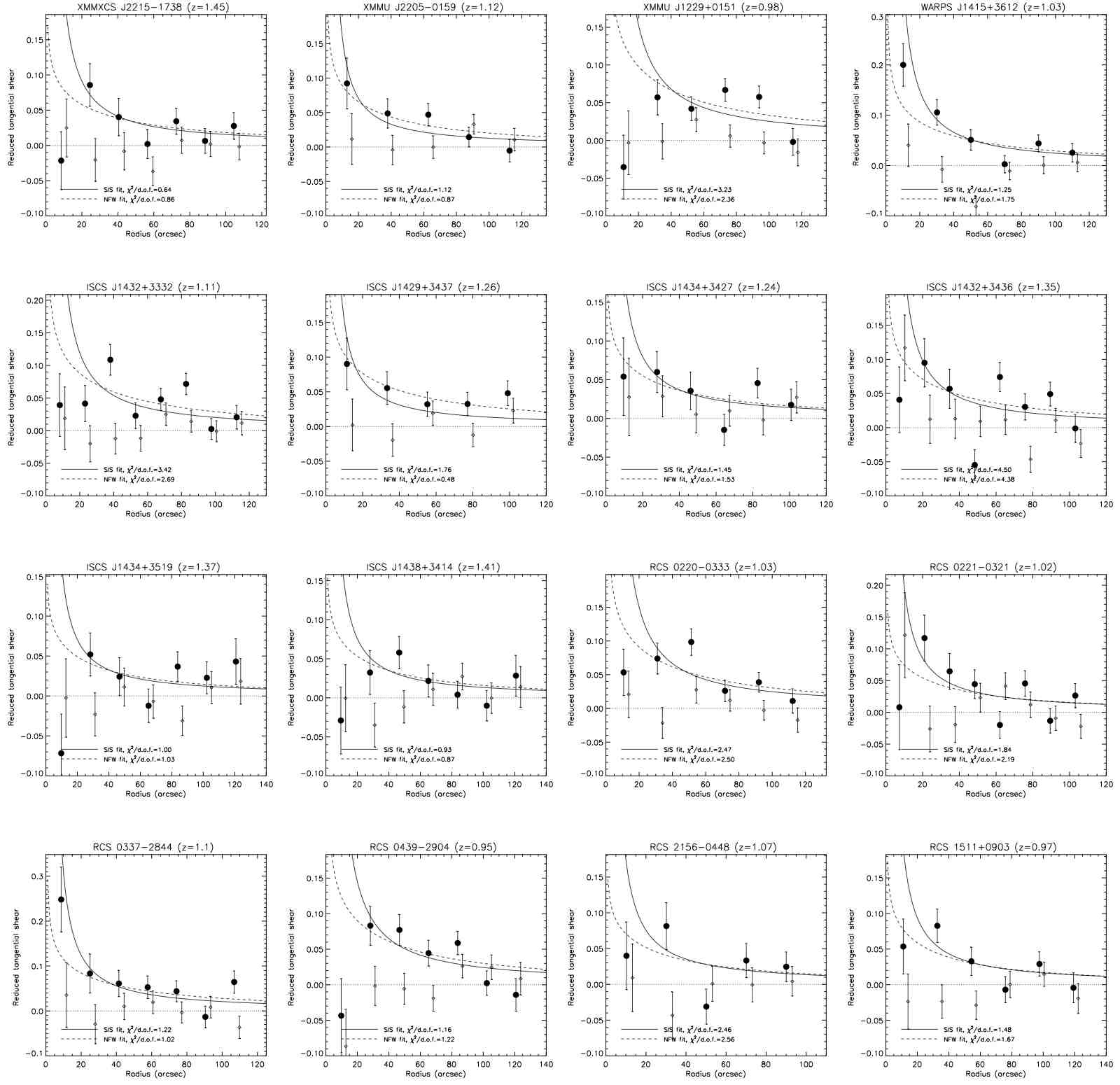}
\caption{Reduced tangential shears for 22 high-redshift clusters. Filled and open circles represent
the tangential shear and 45 $\degr$ rotation test results, respectively.
\label{fig_tan_shear1}}
\end{figure}

\begin{figure}
\plotone{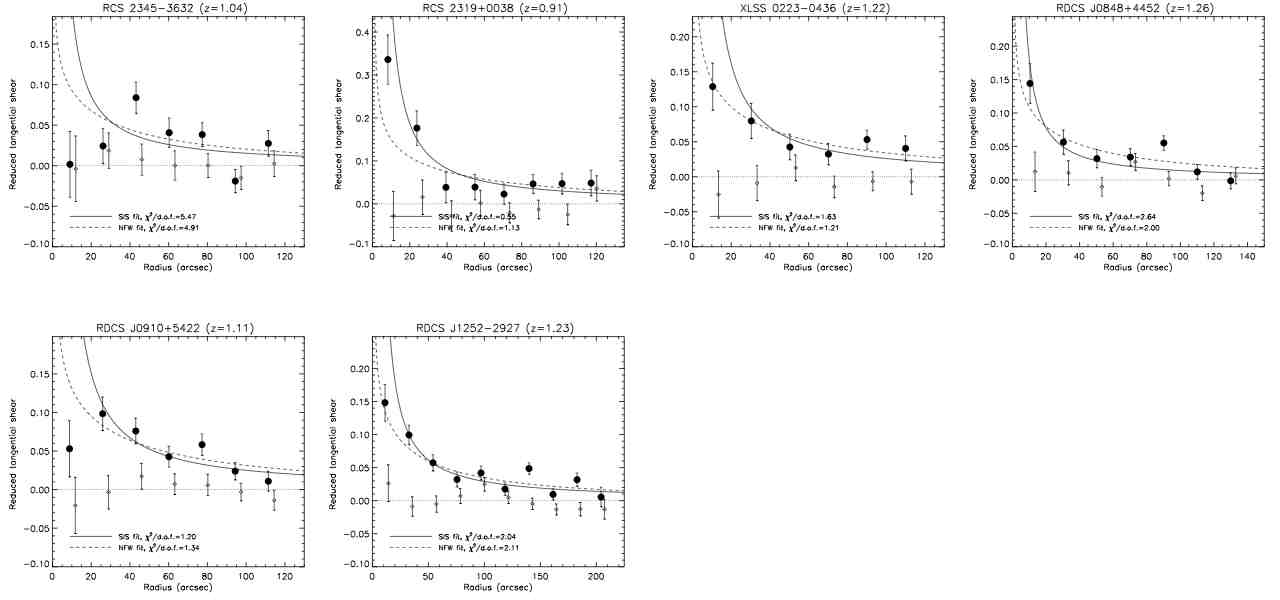}
\caption{Reduced Tangential shears - continued from Figure~\ref{fig_tan_shear1}.
\label{fig_tan_shear2}}
\end{figure}

We employ two kinds of halo models: singular isothermal sphere (SIS) and Navarro-Frenk-White (NFW; Navarro et al. 1997) profiles.
Although the SIS profile is considered inappropriate at large radii (many lines of evidence suggest the mass density drops faster than $1/r^2$), 
the resulting parameter can be conveniently translated to the Einstein radius $\theta_E$ or the velocity dispersion $\sigma_v$ 
of the system. We use this predicted velocity dispersion to compare with the dynamical value. When we fit NFW profiles, it is assumed that the cluster virial mass $M_{200}$ (
the total mass at the radius inside of which the mean density is 200 times the critical density of the universe at the cluster redshift)
is tied to the concentration $c$ via the following relation of Duffy et al. (2008):
\begin{equation}
c=5.71 \left( \frac{M_{200}}{2 \times 10^{12}~h^{-1}~M_{\sun}} \right ) ^{-0.084} (1+z)^{-0.47}.
\end{equation}
\noindent
Therefore, effectively, our NFW model is described by a single parameter. The relation between projected mass density 
and observed shear is simple for SIS with an Einstein radius $r_E$: $\kappa=0.5 r_E / r$ and $g=\kappa / (1-\kappa)$ (in the weak-lensing regime).
For NFW, the relation is rather complicated, and we refer readers to Bartelmann (1996).
We present tangential shears and fitting results for individual clusters in Figure~\ref{fig_tan_shear1}-\ref{fig_tan_shear2}.

In principle, the two-dimensional projected mass distribution $\kappa$ can be obtained by convolving the shear $\gamma$ as follows (Kaiser \& Squires 1993):
\begin{equation}
\kappa (\bvec{x}) = \frac{1}{\pi} \int D^*(\bvec{x}-\bvec{x}^\prime) \gamma (\bvec{x}^\prime) d^2 \bvec{x} \label{eqn_k_of_gamma},
\end{equation}
\noindent
where $D(\bvec{x} ) = - 1/ (x_1 - i x_2 )^2$ is the convolution kernel. However, the direct application of Equation~\ref{eqn_k_of_gamma}
gives rises to some practical problems. First, the ellipticity of individual galaxies should be smoothed, and we do not know the
optimal smoothing scale beforehand. Second, the smoothed galaxy ellipticity gives only an estimate for the reduced shear $g=\gamma/(1-\kappa)$. Third, the relation
$g=\gamma/(1-\kappa)$ is only valid in the $\left | \gamma/ (1-\kappa) \right | < 1$ regime. Fourth, it is not obvious how to weight
shape measurement noise properly via Equation~\ref{eqn_k_of_gamma}. 

Certainly, a more robust two-dimensional mass reconstruction algorithm is required to account for the aforementioned subtleties and
minimize numerous artifacts. Many suggestions are present in the literature (Bridle et al. 1998; Seitz et al. 1998), and the consensus is that a reliable mass reconstruction
should be achieved through iterations. In addition, it is desirable to use the ellipticity of individual galaxies rather than the smoothed values
so as to reveal small-scale, but significant features. 
In the current paper, we reconstruct two-dimensional mass maps using the entropy-regularized, maximum likelihood
code of Jee et al. (2007a). This method of mass reconstruction 
accounts for the aforementioned subtleties and is effective in revealing low-contrast structures.
We present our mass reconstructions in Figure~\ref{fig_xcs2215}-\ref{fig_rdcs1252}.
Results on individual clusters are discussed in \textsection\ref{section_individual}.

\subsection{Redshift Estimation of Source Galaxies }\label{section_redshift_estimation}

Because $\Sigma_c$ scales as $\propto D(z_s)/[D(z_l) D(z_l,z_s) ]$ (eqn. \ref{eqn_sigma_c}), 
the mass density unit rises sharply as the redshift of the lens approaches that of the source.
Thus, in the weak-lensing analysis of high-redshift clusters, small errors in our estimate of the source redshift distribution
result in large errors in the translation of the signal into the physical unit.
For example, a 10\% systematic uncertainty in the effective source redshift leads to $1-2$\% error in mass for a $z=0.2$ object (e.g., ABELL 1689). The 
same amount of uncertainty would give $\sim30$ \% error for the mass of a $z=1.4$ lens 
(e.g., XMM2235-2557)\footnote{The exact value also depends on the depth of the image, which determines how many distant galaxies are resolved. 
Here we assume $\sim 4$ orbit integration with HST/ACS.}.

We estimate the redshift distribution of the source population using the publicly available
Ultra Deep Field (UDF) photometric redshift catalog (Coe et al. 2006). The ultra deep images in six filters
from F435W to F160W provide unparalleled high-quality photometric redshift information well
beyond the limiting magnitude of the cluster images that we analyze here. The catalog has been
extensively used in our previous studies (e.g., Jee et al. 2009), and thus we only briefly summarize the procedure.

First, we bin our source galaxies in 0.5 mag intervals. Next, we randomly draw the galaxies that match
the selection criteria of each bin from the UDF photo-$z$ catalog.
Finally, we combine these simulated catalogs for all bins to create the redshift catalog for the entire source
population. 
Because the UDF is much deeper than the cluster images and also because there is sample variance,
it is important to take into account the difference in number density
in this step. Consequently, we are utilizing the magnitude-color-redshift relation measured in the UDF data, rather than simplistically 
imposing the UDF redshift distribution on our images, which would cause much greater systematics.

As noted in Jee et al. (2009), the total error in $z_{eff}$ due to
he sample variance, the resampling error, and the
difference among the photo-z estimation codes is $\delta z_{eff}\simeq0.06$. 
This causes $\sim11$\% ($\sim3$ \%) uncertainty in mass for a $z\sim1.4$ ($z\sim0.9$) cluster. We include these errors
in our final uncertainty.
The assumption that there exists a single source plane while in reality each source galaxy is at different redshift
also creates a bias in the interpretation of the lensing signal. We correct this bias to first order using the
equation derived by Seitz \& Schneider (1997).

\subsection{Notes on Individual Clusters} \label{section_individual}

We here comment on specific features in the measurement of each cluster. All of the key numbers are summarized in Table 2.

\subsubsection{XMMXCS J2215-1738 ($z=1.46$) }
XMMXCS2215-1738 was the highest redshift cluster spectroscopically confirmed (Stanford et al. 2006)
until the recent discovery of ClG J0218.3-0510 at $z=1.62$ (Papovich et al. 2010).
Assuming no significant point source contamination in the archival 
XMM-$Newton$ data, Stanford et al. (2006) obtained a temperature of $T_X = 7.4_{-1.8}^{+2.7}$ keV
with a bolometric luminosity of $L_X=4.4_{-0.6}^{+0.8} \times 10^{44} \mbox{ergs}~\mbox{s}^{-1}$.
Hilton et al. (2010) reported significantly lower values $T_X=4.1_{-0.9}^{+0.6}$ keV and 
$L_X=2.92_{-0.35}^{+0.24}\times10^{44} \mbox{ergs}~\mbox{s}^{-1}$ based on
a joint analysis of Chandra and XMM-$Newton$ data, which enables them to
remove the point source contamination from the diffuse component.
Assuming an isothermal $\beta$ model with $\beta=0.63$ and $r_c=52$ kpc, 
the temperature $T_X=4.1_{-0.9}^{+0.6}$ keV translates into $M_{200}=2.0_{-0.6}^{+0.5}\times10^{14}~M_{\sun}$.

\begin{figure}
\plotone{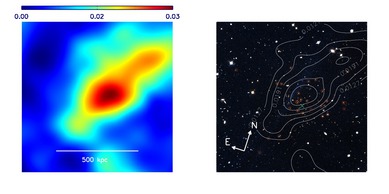}
\caption{XMMXCS J2215-1738 mass map. The left panel displays the mass reconstruction, whose color represents
the mass density. The mass-sheet degeneracy is not broken here, and thus the scale is arbitrary.
On the right panel, we overlay the mass contours on the pseudo-color composite created by
combining the $i_{775}$ and $z_{850}$ images. The green ``X'' symbol marks the location of the X-ray peak in
the $Chandra$ image. Spectroscopically confirmed members are encircled in red.
\label{fig_xcs2215}}
\end{figure}

We initially considered a weak-lensing study of XMMXCS2215-1738 to be highly challenging
because of the low X-ray mass, high redshift, and relatively shallow depth of the F775W image, where
we measure galaxy shapes. However, the cluster is clearly detected in our two-dimensional mass reconstruction, which
shows the mass centroid in good spatial agreement with the cluster members and the X-ray centroid (Figure~\ref{fig_xcs2215}).
The weak-lensing mass of the cluster is $M_{200} = 4.3_{-1.7}^{+3.0}\times10^{14}~M_{\sun}$, a factor of two higher than the X-ray estimate (however,
both results are marginally consistent). 

The predicted velocity dispersion $942_{-126}^{+111}~\mbox{km~s}^{-1}$ is consistent with the
spectroscopically measured value $710\pm110~\mbox{km~s}^{-1}$ derived from 31 members within $R_{200}$ (Hilton et a. 2010);
interestingly, if all 44 members were used, the velocity dispersion increases to $890\pm110~\mbox{km~s}^{-1}$ giving
a better agreement with the lensing result.

\subsubsection{XMMU J2205-0159 ($z=1.12$) }

\begin{figure}
\plotone{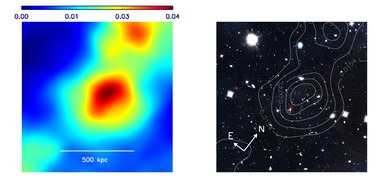}
\caption{Same as Figure~\ref{fig_xcs2215} but for XMMU J2205-0159. 
\label{fig_xmm2205}}
\end{figure}

XMMU2205-0159 was discovered in the XDCP (Boehringer et al. 2006) survey in the archival image of
QSO B2202-0209 at z=1.77.
The mass centroid is offset north of the BCG $(\alpha,\delta)\simeq$(22:05:50.7,-01:59:30) by $\sim17\arcsec$ 
(Figure~\ref{fig_xmm2205}). We derive a weak-lensing mass of $M_{200}=3.0_{-1.0}^{+1.6}\times10^{14}~M_{\sun}$, making the cluster among 
the least massive clusters in our sample. Neither a dynamical nor X-ray study of the cluster 
has been published to date.

\subsubsection{XMMU J1229+0151 ($z=0.98$) }

\begin{figure}
\plotone{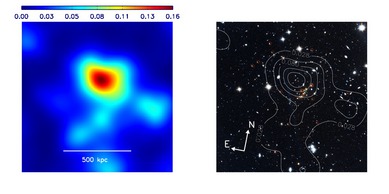}
\caption{Same as Figure~\ref{fig_xcs2215} but for XMMU J1229+0151. 
\label{fig_xmm1229}}
\end{figure}

XMMU J1229+0151 was serendipitously detected in the field targeted for the well-known quasar 3C 273. 
The cluster is hot and X-ray luminous with $T_X=6.4_{-0.6}^{+0.7}$ keV and $L_X=3\times10^{44}~\mbox{erg~s}^{-1}$ (Santos et al. 2009).
Our lensing analysis predicts a velocity dispersion of $867_{-69}^{+64}~\mbox{km~s}^{-1}$ and gives a virial mass of
$M_{200}=5.3_{-1.2}^{+1.7}\times10^{14}~M_{\sun}$. This lensing mass agrees nicely with
the X-ray mass $M_{200}=5.7_{-0.8}^{+1.0}\times10^{14}~M_{\sun}$ when an isothermal $\beta$ model
with $\beta=0.7$ and $r_c=250$ kpc is assumed; no measurement of the X-ray surface brightness has been reported yet.
However, the velocity dispersion $683\pm62~\mbox{km~s}^{-1}$ derived from 27 spectroscopic members (Santos et al. 2009) is
significantly lower than the lensing prediction $867_{-69}^{+64}~\mbox{km~s}^{-1}$.
Our convergence map shows a strong peak $\sim10\arcsec$ northeast of the X-ray peak and the cluster galaxies (Figure~\ref{fig_xmm1229}). 

\subsubsection{WARPS J1415+3612 ($z=1.03$) }

The cluster WARPS J1415+3612 was discovered (Perlman et al. 2002) in the first phase of
the Wide Angle $ROSAT$ Pointed Survey (WARPS). The ICM temperature of the cluster has been
measured from the $Chandra$ ($5.59\pm0.84$ keV; Allen et al. 2008) and XMM-$Newton$
($5.7_{-0.7}^{+1.2}$ keV; Maughan et al. 2006) data analysis. 
The cluster has been considered relaxed because of its symmetric X-ray emission
(Maughan et al. 2006; Allen et al. 2008). Our weak-lensing mass map of the cluster also
does not show any significant substructure, adding an additional support to the
hypothesis. Both virial masses derived from X-ray and lensing 
($M_{200}=4.6_{-0.8}^{+1.5}\times10^{14}~M_{\sun}$ and $4.7_{-1.4}^{+2.0}\times10^{14}~M_{\sun}$, respectively  )   are
in excellent agreement.

\begin{figure}
\plotone{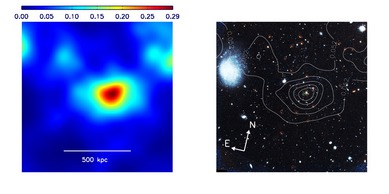}
\caption{Same as Figure~\ref{fig_xcs2215} but for WARPS J1415+3612.
\label{fig_warps1415}}
\end{figure}

The ACS image of the cluster shows a strongly lensed arc at $z=3.898$. The source
is a Ly-$\alpha$ emitter, and is located $\sim7\arcsec$ away from the BCG (Huang et al. 2009).
When we adopt this distance as the Einstein radius at $z=3.898$, the value is
consistent with the weak-lensing prediction $\theta_E(z=3.898)=8\arcsec\pm2\arcsec$ derived
from the NFW fitting result; note also that the location of the weak-lensing mass peak is in
good agreement with that of the BCG (Figure~\ref{fig_warps1415}).

\subsubsection{ISCS J1432+3332 ($z=1.11$) }

\begin{figure}
\plotone{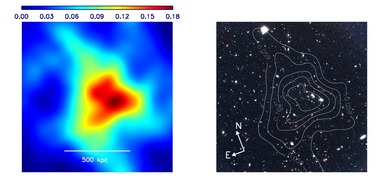}
\caption{Same as Figure~\ref{fig_xcs2215} but for ISCS J1432+3332.
\label{fig_iscs1432+3332}}
\end{figure}

The ISCS J1432+3332 cluster was reported by Elston et al. (2006) as one of the first spectroscopically confirmed $z>1$ clusters
detected through the joint use of 
the FLAMINGOS Extragalactic Survey (FLAMEX; Elston et al. 2006), the NOAO Deep Wide-Field Survey (NDWFS; Brown et al. 2003), and 
the Spitzer IRAC Shallow Survey (ISCS; Eisenhardt et al. 2004) data sets. They estimated from 8 spectroscopic members that the velocity
dispersion of the cluster is $920\pm230 \mbox{km}~\mbox{s}^{-1}$.
Eisenhardt et al. (2008) reported a refined measurement of $\sigma_v=734~\mbox{km}~\mbox{s}^{-1}$ using
23 redshifts; the authors did not quote an uncertainty and we estimate $\delta\sigma_v\sim115~\mbox{km}~\mbox{s}^{-1}$.
The virial mass is estimated to be 
$M_{200}=4.9_{-1.2}^{+1.6}\times10^{14}~M_{\sun}$. The mass contours are well traced by the cluster galaxies
(Figure~\ref{fig_iscs1432+3332}).

\subsubsection{ISCS J1429+3437 ($z=1.26$) }

ISCS J1429+3437 was identified photometrically with the combined use of the 
ISCS and NDWFS data sets (Eisenhardt et al. 2008). From the 9 spectroscopic
members of Eisenhardt et al. (2008), we calculate the velocity dispersion
to be $767\pm295~\mbox{km}~\mbox{s}^{-1}$, which
agrees well with the predicted velocity dispersion 
$732_{-78}^{+70}~\mbox{km~s}^{-1}$. For the virial mass, we
obtain $M_{200}=5.4_{-1.6}^{+2.4}\times10^{14}~M_{\sun}$.

\begin{figure}
\plotone{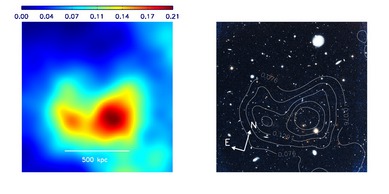}
\caption{Same as Figure~\ref{fig_xcs2215} but for ISCS J1429+3437.
\label{fig_iscs1429+3437}}
\end{figure}

\begin{figure}
\plotone{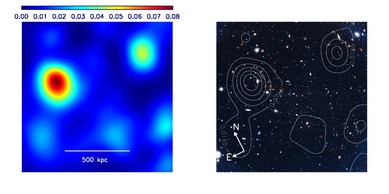}
\caption{Same as Figure~\ref{fig_xcs2215} but for ISCS J1434+3427.
\label{fig_iscs1434+3427}}
\end{figure}

\begin{figure}
\plotone{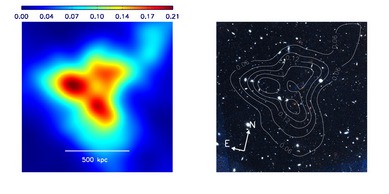}
\caption{Same as Figure~\ref{fig_xcs2215} but for ISCS J1432+3436.
\label{fig_iscs1432+3436}}
\end{figure}

We detect two mass clumps within the ACS field of the cluster observation (Figure~\ref{fig_iscs1429+3437}). The western
mass peak is much stronger and spatially correlated with the cluster galaxy candidates.
The much weaker eastern clump does not have any compact galaxy distribution. Nevertheless, the
global east-west elongation of the mass distribution seems to follow the early-type galaxies in the
field whose colors are consistent with that of the red-sequence at the redshift of the cluster.

\subsubsection{ISCS J1434+3427 ($z=1.24$) }
The ISCS J1434+3427 cluster, discovered in the ISCS and NDWFS survey, is
reported to possess a pronounced filamentary structure (Brodwin et al.  2006).
Although our mass reconstruction reveals a significant mass peak 10 $\arcsec$ east of the compact
galaxy distribution, the current $i_{775}$ image is not deep enough ($\sim2685$ s)
to study the substructure of this high-redshift cluster in detail.
Because the cluster is located near the edge of the ACS field, we need to assume
an azimuthal symmetry at large radius to obtain a mass estimate. This may substantially
bias our measurement if the cluster departs significantly from the assumed axisymmetry.
Nevertheless, we note that the predicted velocity dispersion $770_{-133}^{+113}~\mbox{km}~\mbox{s}^{-1}$ by our lensing analysis is consistent with
the dynamical velocity dispersion $863\pm170~\mbox{km}~\mbox{s}^{-1}$ that are derived from 11 cluster members (Meyers et al.  2011).

\subsubsection{ISCS J1432+3436 ($z=1.35$) }

The cluster was discovered in the ISCS and NDWFS survey, and about 8
members have been spectroscopically confirmed (Eisenhardt et al. 2008). 
The cluster detection is very strong in our mass reconstruction despite
both the high-redshift and the shallowness (2235 s) of the $i_{755}$ image.
We estimate the virial mass to be $M_{200}=5.3_{-1.7}^{+2.6}\times10^{14}~M_{\sun}$.
The velocity dispersion derived from 8 members is $807\pm340~\mbox{km}~\mbox{s}^{-1}$, which 
agrees well with the lensing prediction $912_{-102}^{+92}~\mbox{km}~\mbox{s}^{-1}$.

\subsubsection{ISCS J1434+3519 ($z=1.37$) }

\begin{figure}
\plotone{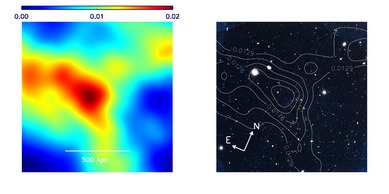}
\caption{Same as Figure~\ref{fig_xcs2215} but for ISCS J1434+3519.
\label{fig_iscs1434+3519}}
\end{figure}

\begin{figure}
\plotone{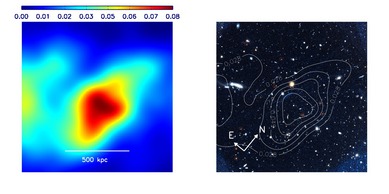}
\caption{Same as Figure~\ref{fig_xcs2215} but for ISCS J1438+3414.
\label{fig_iscs1438}}
\end{figure}

\begin{figure}
\plotone{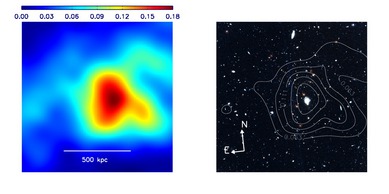}
\caption{Same as Figure~\ref{fig_xcs2215} but for RCS 0220-0333.
\label{fig_rcs0220}}
\end{figure}

\begin{figure}
\plotone{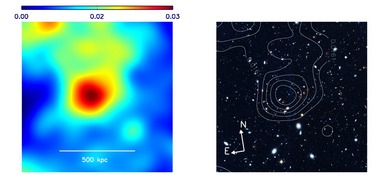}
\caption{Same as Figure~\ref{fig_xcs2215} but for RCS 0221-0321.
\label{fig_rcs0221}}
\end{figure}

\begin{figure}
\plotone{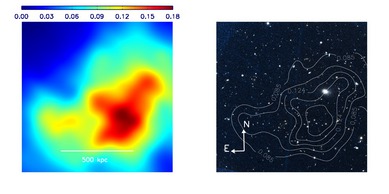}
\caption{Same as Figure~\ref{fig_xcs2215} but for RCS 0337-2844.
\label{fig_rcs0337}}
\end{figure}

\begin{figure}
\plotone{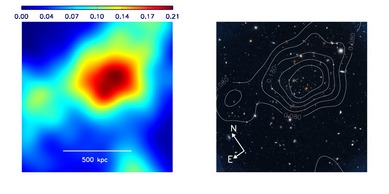}
\caption{Same as Figure~\ref{fig_xcs2215} but for RCS 0439-2904.
\label{fig_rcs0439}}
\end{figure}

\begin{figure}
\plotone{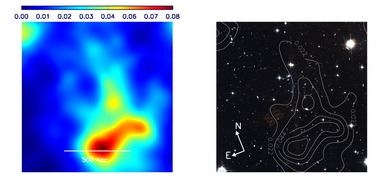}
\caption{Same as Figure~\ref{fig_xcs2215} but for RCS 2156-0448.
\label{fig_rcs2156}}
\end{figure}

\begin{figure}
\plotone{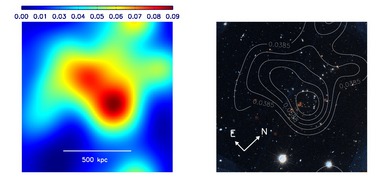}
\caption{Same as Figure~\ref{fig_xcs2215} but for RCS 1511+0903.
\label{fig_rcs1511}}
\end{figure}

Eisenhardt et al. (2008) report that five members have been spectroscopically
confirmed. Our two-dimensional mass reconstruction (Figure~\ref{fig_iscs1434+3519}) reveals a weak detection of convergence peak
near the cluster center defined by Eisenhardt et al. (2008) using photometric redshifts.
The $15\arcsec$ offset may be attributed to statistical noise because of the high redshift, the low mass, and the
insufficient depth (1920 s) of the $i_{775}$ image. The virial mass is determined to be 
$M_{200}=2.8_{-1.4}^{+2.9}\times10^{14}~M_{\sun}$. 

\subsubsection{ISCS J1438+3414 ($z=1.41$) }

The discovery of ISCS J1438+3414 is reported by Stanford et al. (2005), who
confirmed five spectroscopic members. Brodwin et al. (2011) quote a dynamical velocity
dispersion of $757_{-203}^{+247}~\mbox{km}~\mbox{s}^{-1}$ from a total of 11 cluster members, which
can be converted to $M_{200}=2.3^{+2.4}_{-2.1} \times 10^{14}~M_{\sun}$ using the mass-dispersion relation from Evrard et al. (2008).
These values are consistent with those from our lensing analysis, which give $833_{-150}^{+127}~\mbox{km}~\mbox{s}^{-1}$ and a virial mass of
$M_{200}=3.1_{-1.4}^{+2.6}\times10^{14}~M_{\sun}$.
Andreon et al. (2011) measured an X-ray temperature $T_X=4.9_{-1.6}^{+3.4}$ keV 
from the relatively deep ($\sim150$ ks) chandra data of the cluster. This temperature is consistent
with the measurement ($3.3_{-1.0}^{+1.9}$ keV) of Brodwin et al. (2011) obtained from the same data. 
Assuming $\beta=0.7$ and $rc\sim100$ kpc, we 
convert  $T_X=4.9_{-1.6}^{+3.4}$ keV to $M^X_{200}=3.2_{-1.4}^{+3.8}~M_{\sun}$, which also agrees well with our lensing mass. 
The centroid of the diffuse X-ray emission is in good spatial
agreement with that of the weak-lensing mass.

\subsubsection{RCS 0220-0333 ($z=1.03$) }

The clusters with a prefix RCS hereafter were mostly discovered
in the Red-sequence Cluster Survey-I (Gladders \& Yee 2005); the exception is RCS 1511+0903, which was discovered in 
the Red-sequence Cluster Survey-II (Gilbank et al. 2011).
RCS 0220-0333 is an optically rich cluster at $z=1.03$.
In the ACS pseudo-color composite image, the early-type galaxies with $i_{775}-z_{850}\sim0.9$
appear to form a North-South filamentary structure.
The cluster is strongly detected in lensing. The mass centroid lies close to
the cluster member with pronounced strong lensing features.
However, the mass distribution does not indicate the North-South elongation
seen in the cluster galaxies. 
We obtain an virial mass of $M_{200}=4.8_{-1.3}^{+1.8}\times10^{14}~M_{\sun}$ with
a predicted velocity dispersion of $881_{-74}^{+68}~\mbox{km}~\mbox{s}^{-1}$. 

\subsubsection{RCS 0221-0321 ($z=1.02$) }

This optically rich galaxy cluster RCS 0221-0321 is
clearly visible in our mass reconstruction. The mass map show
a significant mass clump aligned with the optical center (e.g.,  see the isodensity contours for red galaxies from Andreon et al. 2008). 
However, this cluster is found to
be one of the least massive clusters in our sample. We
estimate the virial mass to be $M_{200}=1.8_{-0.7}^{+1.3}\times10^{14}~M_{\sun}$.
The predicted velocity dispersion $699_{-94}^{+83}~\mbox{km}~\mbox{s}^{-1}$ agrees
well with the dynamical measurement $710\pm150~\mbox{km}~\mbox{s}^{-1}$ based on
21 spectroscopic members (Andreon et al. 2008).

\subsubsection{RCS 0337-2844 ($z=1.10$) }

RCS 0337-2844 is clearly visible in our two-dimensional mass map, which
shows a good mass-galaxy correlation.
The exposure time of the $i_{755}$ image of this cluster 
is the shortest among the clusters studied here, which indicates that this strong
weak-lensing signal is due to a significant mass.
We estimate $M_{200}=4.9_{-1.7}^{+2.8}\times10^{14}~M_{\sun}$. No spectroscopic
data have been published to date.

\subsubsection{RCS 0439-2904 ($z=0.95$) }

Barrientos et al. (2004) reported that RCS 0439-2904 is an optically
rich cluster ($\sim9~\sigma$ in the $K_s$ image) at a high redshift. 
Cain et al. (2008) measured the X-ray temperature
of the cluster from the $200$ks Chandra to be $T_x=1.5_{-0.4}^{+1.0}$ keV, which
is significantly lower than what the optical richness suggested.
The single isothermal $\beta$ model gives $M_{200}=4.6_{-1.7}^{+6.0}\times10^{13}~M_{\sun}$.
On the other hand, the mass-richness relation implies a much higher mass by at least an order of magnitude.
Because their velocity histogram show that there are multiple components, 
Cain et al. (2008) interpreted this large discrepancy as indicating
the presence of two clusters along the line-of-sight direction. They also claim
that the unexpectedly high gas fraction obtained when a single component is assumed
supports the two-component interpretation.

The line-of-sight hypothesis makes our lensing study of the cluster interesting.
If there is indeed a line-of-sight superposition of two clusters, the mass estimate
from lensing should give the sum of the two clusters as opposed to the single component
X-ray model. Our analysis shows that the cluster virial mass is
$M_{200}=4.3_{-1.2}^{+1.7}\times10^{14}~M_{\sun}$, nearly a factor of 10 higher than the
X-ray prediction as was also indicated by the mass-richness relation.
Interestingly, our predicted velocity dispersion $831_{-74}^{+68}~\mbox{km}~\mbox{s}^{-1}$
is consistent with the dynamical measurement $1080\pm320~\mbox{km}~\mbox{s}^{-1}$ of Cain et al. (2008).
The lensing mass is consistent with the mass-richness relation if the cluster
is assumed to consist of two clusters; the new lensing mass shifts the Model II data
point of Figure 3 of Cain et al. (2008) upward in such a way that the new point is
well bracketed by the $1-\sigma$ scatter of the mass-richness relation of Blindert (2006).

\subsubsection{RCS 2156-0448 ($z=1.07$) }

The cluster RCS 2156-0448 was reported as an optically rich cluster with a strong
lensing arc candidate by Gladders et al. (2003). If confirmed, the
presence of this arc suggests that the cluster is massive as
indicated by its optical richness. However, Gladders et al. (2003)
commented that this candidate should be considered the least likely among
their secondary sample.

Our weak-lensing mass reconstruction reveals only a weak convergence 
peak near the arc candidate. Instead, the mass map shows a more significant
mass peak $\sim70\arcsec$ south of the assumed cluster center.
Our visual inspection of the ACS image shows that on this location there seems to be also
early-type galaxies whose colors are consistent with that of the cluster red-sequence.

Nevertheless, one must be reminded that the combined $(i_{775}+z_{850})$ ACS image of the cluster is the shallowest among our 22 cluster sample, which
gives the lowest number density of background galaxies ($\sim50~\mbox{arcmin}^{-2}$). 
Hence, the mass-galaxy comparison should await a future analysis with deeper images.
Our mass presented here is estimated by placing the center on the strong-lensing candidate of Gladders et al. (2003).

\subsubsection{RCS 1511+0903 ($z=0.97$) }

The ACS image of RCS 1511+0903 shows a compact distribution of early-type
galaxies whose colors are consistent with the cluster redshift $z=0.97$.
The cluster is clearly visible in our mass reconstruction, which reveals
a strong mass peak at the location of the cluster galaxies. 
Based on 9 redshifts, Meyers et al. (2011) estimate the velocity dispersion of the cluster to be
$717\pm208~\mbox{km}~\mbox{s}^{-1}$, which is consistent with our lensing prediction
$699_{-109}^{+94}~\mbox{km}~\mbox{s}^{-1}$. The virial mass of the cluster from our lensing
analysis is $M_{200}=1.9_{-0.8}^{+1.4}\times10^{14}~M_{\sun}$.

\subsubsection{RCS 2345-3632 ($z=1.04$) }

\begin{figure}
\plotone{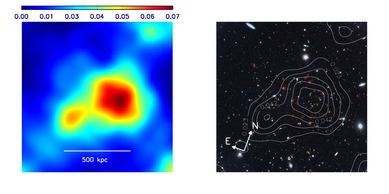}
\caption{Same as Figure~\ref{fig_xcs2215} but for RCS 2345-3632.
\label{fig_rcs2345}}
\end{figure}

\begin{figure}
\plotone{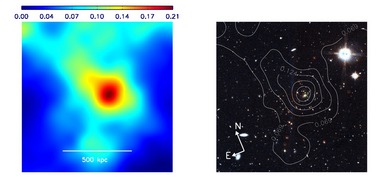}
\caption{Same as Figure~\ref{fig_xcs2215} but for RCS 2319+0038.
\label{fig_rcs2319}}
\end{figure}

\begin{figure}
\plotone{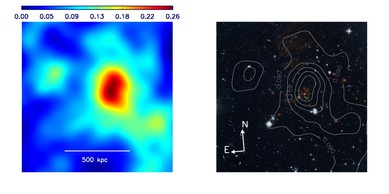}
\caption{Same as Figure~\ref{fig_xcs2215} but for XLSS J0223-0436.
\label{fig_xlss0223}}
\end{figure}

RCS 2345-3632 is an optically rich cluster at z=1.04 with 23 spectroscopically confirmed cluster galaxies.
The dynamical velocity dispersion $670\pm190~\mbox{km}~\mbox{s}^{-1}$ is consistent with
the lensing prediction $684_{-79}^{+71}~\mbox{km}~\mbox{s}^{-1}$.
The two-dimensional mass map agrees well with the cluster galaxy distribution (Figure~\ref{fig_rcs2345}).
The relatively deep ACS image allows us to utilize $\sim103$ background galaxies per sq. arcmin, and thus
the resulting mass reconstruction should be considered most reliable among the 8 RCS clusters presented
in this paper. We estimate that the virial mass of the cluster is $M_{200}=2.4_{-0.7}^{+1.1}\times10^{14}~M_{\sun}$.

\subsubsection{RCS 2319+0038 ($z=0.91$) }

RCS 2319+0038 is  a strong-lensing cluster 
showing at least two spectacular arcs (Gladders et al.  2003). 
The recent ACS images of the cluster
reveal more strong-lensing features (at least 7 tangential and
2 radial arc candidates). 
Gilbank et al. (2008)
discovered that the cluster is in fact part of a supercluster containing 
two other clusters RCS 2319+0030 and RCS 2320+0033. All three
clusters are well detected in the $Chandra$ X-ray observations with
similar gas temperatures $\sim6$ keV (Hicks et al. 2008).
RCS 2319+0038 is separated from RCS 2320+0033 and RCS 2319+0030 by $\sim5\arcmin$
and $\sim7\arcmin$, respectively, and therefore our lensing analysis
based on the ACS image covering the central $3\arcmin\times3\arcmin$ region of
RCS 2319+0038 is not likely to be affected by these two other structures.
Gilbank et al. (2008) reported a velocity dispersion of $990\pm240~\mbox{km}~\mbox{s}^{-1}$,
which agrees well with the current lensing prediction $898_{-71}^{+67}~\mbox{km}~\mbox{s}^{-1}$.
Our convergence map reveals a single strong mass peak at the center of the strong-lensing
system without any significant substructures. Both this mass map and the tangential shear profile
suggest that the cluster is relaxed. The virial mass derived from the
X-ray measurements of Hicks et al. (2008) ($T_X=6.2_{-0.8}^{+0.9}$ keV, $r_c=100$ kpc, and $\beta=0.65$)
is $M_{200}=5.4_{-1.0}^{+1.2}\times10^{14}~M_{\sun}$, again in good agreement with
the lensing result $M_{200}=5.8_{-1.6}^{+2.3}\times10^{14}~M_{\sun}$.

\subsubsection{XLSS 0223-0436 ($z=1.22$) }

XLSS 0223-0436 was discovered in the XMM Large Scale Structure (LSS) survey
(Pierre et al. 2004; Andreon et al. 2005), and Bremer et al. (2006) presented detailed analysis
of the X-ray and optical/near-IR data of the cluster. They estimated
an X-ray temperature of 3.8 keV with a $1-\sigma$ lower limit of 1.9 keV and
an unconstrained upper limit.
This X-ray temperature 3.8 keV is somewhat lower than what one predicts
from the dynamical velocity dispersion $799\pm129~\mbox{km}~\mbox{s}^{-1}$ (based on 24 redshfits). Because
Bremer et al. (2006) report that the existing XMM-Newton data do not
constrain the upper limit of their temperature, this cluster is excluded in our investigation of the lensing mass-temperature relation.
Our lensing analysis predicts $\sigma_v=1011_{-79}^{+73}~\mbox{km}~\mbox{s}^{-1}$, which is
even higher than the dynamical measurement. 
We estimate that the virial mass of the cluster is
$M_{200}=7.4_{-1.8}^{+2.5}\times10^{14}~M_{\sun}$.

Bremer et al. (2006) reported that the optical image of the cluster shows a compact
distribution of $>12$ galaxies within a 125 kpc radius centered on the X-ray peak.
The location of our weak-lensing mass peak agrees well with the X-ray and
the optical centers.

\subsubsection{RDCS J0849+4452 ($z=1.26$) }

\begin{figure}
\plotone{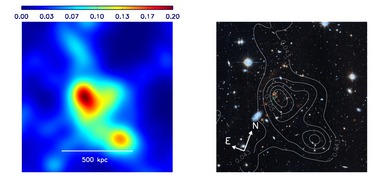}
\caption{Same as Figure~\ref{fig_xcs2215} but for RDCS J0849+4452.
\label{fig_rdcs0848}}
\end{figure}

\begin{figure}
\plotone{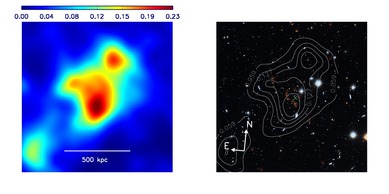}
\caption{Same as Figure~\ref{fig_xcs2215} but for RDCS J0910+5422.
\label{fig_rdcs0910}}
\end{figure}

\begin{figure}
\plotone{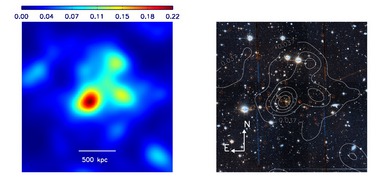}
\caption{Same as Figure~\ref{fig_xcs2215} but for RDCS J1252-2927.
\label{fig_rdcs1252}}
\end{figure}

RDCS J0849+4452 was discovered in the ROSAT Deep Cluster Survey (RDCS;  Rosati et al. 1998), 
and followup near-IR imaging showed an excess of red (1.8 $< J-K <$ 2.1) 
galaxies around the peak of the X-ray emission. Cluster galaxies were spectroscopically
confirmed using the Keck telescope (Rosati et al. 1999). Stanford et al. (2001)
measured an X-ray temperature $T_x = 5.8_{-1.7}^{+2.8}$ keV from the $Chandra$ data analysis;
based on a newer calibration data, Ettori et al. (2009) reported $T_x = 5.2\pm1.3$ keV.
The cluster was one of the ACS GTO high-redshift cluster targets, and
the color magnitude relation and
the weak-lensing analysis have been presented by Mei et al. (2006a) and Jee et al. (2006),
respectively.
Jee et al. (2006) quoted a projected mass of $M_{proj}=(2.1\pm0.7)\times10^{14}~M_{\sun}$
within a 0.5 Mpc aperture radius, which is consistent with the projected X-ray mass
$\sim2.3\times10^{14}~M_{\sun}$, which was obtained from their re-analysis
of the $Chandra$ data.
In the current paper, we perform a weak-lensing analysis using a
slightly deeper set of the ACS images, the result of stacking
both the GTO data (Prop. ID 9919) and the high-z supernova search data (Prop. ID 10496). 
The new mass is in good agreement with the Jee et al. (2006) result;
in the current paper we quote a virial mass $M_{200}=4.4_{-0.9}^{+1.1}\times10^{14}~M_{\sun}$ ($r_{200}=0.98$)
rather than a projected mass. The cluster shows a strong-lensing features around the BCG.
The two-dimensional mass map (Figure~\ref{fig_rdcs0848}) reveals two significant mass clumps. The stronger one
coincides with the centers of the X-ray emission and the optical center while the weaker seems to
be associated with a foreground group. Both clumps were also shown in the mass reconstruction of Jee et al. (2006).

\subsubsection{RDCS J0910+5422 ($z=1.11$) }

Discovered in the RDCS, the cluster
was confirmed with near-IR and spectroscopic observations by Stanford et al. (2002).
Ettori et al. (2009) measured an X-ray temperature of $6.4\pm1.4$ keV
from 200 ks $Chandra$ data, slightly
lower than the measurement ($7.2_{-1.4}^{+2.2}$ keV) by Stanford et al. (2002).
Combining $6.4\pm1.4$ keV with $r_c=147$ kpc and $\beta=0.843$, we
obtain $M_{200}=7.7_{-2.2}^{+3.2}\times10^{14}~M_{\sun}$ assuming an isothermal $\beta$ profile.
Our weak-lensing analysis yields a cluster mass $M_{200}=5.0_{-1.0}^{+1.2}\times10^{14}~M_{\sun}$ ($r_{200}=1.07$ Mpc), 
slightly lower than, but statistically consistent with this X-ray result.
The velocity dispersion estimated from 25 redshifts by Mei et al. (2006) is $675\pm190~\mbox{km}~\mbox{s}^{-1}$, lower than
our lensing prediction of $895_{-51}^{+48}~\mbox{km}~\mbox{s}^{-1}$ and  the X-ray mass.
Our mass reconstruction reveals two significant peaks. The stronger one is in good spatial agreement
with the optical and X-ray centers whereas no apparent red-sequence is found at the location of the secondary peak.

\subsubsection{RDCS J1252-2927 ($z=1.24$) }
RDCS J1252-2927 was confirmed as a cluster at $z=1.24$ based on
an extensive spectroscopic campaign using the Very Large
Telescope (VLT) (Lidman et al. 2004). Rosati et al. (2004)
presented the first X-ray analysis of the cluster based on both deep
$Chandra$ and $XMM-Newton$ observations, which
gives an X-ray temperature of $T_X=6.0_{-0.5}^{+0.7}$ keV.
A revised temperature of $T_X=6.5\pm0.5$ keV is reported in 
Lombardi et al. (2005) after the application of the new
calibration of the $Chandra$ instrument. The most recent analysis by Ettori et al. (2009)
gives $T_X=7.6\pm1.2$ keV.

Assuming the structural parameters $r_c=79$ kpc and
$\beta=0.53$, we translate the temperature $T_x=7.6\pm1.2$ keV
into a virial mass $M_{200}=(4.4\pm1.0)\times10^{14}~M_{\sun}$
or a projected mass at $r=1$ Mpc $M_{proj}=7.1\pm1.1\times10^{14}~M_{\sun}$.
Using ACS data, Lombardi et al. (2005) quote a projected weak-lensing mass of $M(<1\mbox{Mpc})=(8.0\pm1.3)\times10^{14}~M_{\sun}$, which
is higher than the X-ray result by $\sim1\sigma$.
Our lensing analysis based on deeper ACS images gives a virial mass of $M_{200}=6.8_{-1.0}^{+1.2}\times10^{14}~M_{\sun}$ or
a projected mass of $M(<1\mbox{Mpc})=(8.4\pm1.2)\times10^{14}~M_{\sun}$, which is
consistent with the result of Lombardi et al. (2005).
The current
lensing analysis predicts $\sigma_v=957_{-48}^{+45}~\mbox{km}~\mbox{s}^{-1}$, higher than the dynamical measurement 
$\sigma_v=747_{-84}^{+74}~\mbox{km}~\mbox{s}^{-1}$ that Demarco et al. (2007) obtained from 38 redshifts.
Lombardi et al. (2005) discussed the possibility of a systematic overestimation of the mass in lensing because
of a line-of-sight contamination. Demarco et al. (2007) reported that there is a possible group centered at $z=0.74$ composed
of 33 members in the range $0.70<z<0.79$. The projected distribution, however, is not compact, but is mostly scattered across
the field. In addition, the deep ($\sim 190$ ks) $Chandra$ data do not hint at the presence of any significant intervening structure.

Our mass reconstruction reveals three significant mass peaks. The strongest one coincides
with the optical and the X-ray centers. The other two weaker clumps do not correlate well with the
red-sequence of the cluster nor the group at $z=0.74$. We note that these two mass clumps are
also clearly seen in Figure 9 of Lombardi et al. (2005).

\section{COMPARISON WITH OTHER CLUSTER PROPERTIES} \label{section_comparison}
The weak-lensing masses presented in the current work are used to study
mass versus observable relations in the high-redshift universe. A comparable study but for
a lower redshift sample was carried out by Hoekstra (2007) for 20 X-ray luminous clusters. Our study
enables us to examine if the mass versus mass-observable
relation holds across a wide range of redshifts. 

As many clusters in our sample are the results of relatively recent discoveries,
a large fraction of them still lack information on their velocity dispersion and X-ray properties.
In the subsequent analysis, we enlarge the current sample by adding five more high-redshift clusters in our
previous lensing studies: XMMU J2235.3-2557 $z=1.4$ (Jee et al. 2009), CL J1226+3332 $z=0.89$ (Jee \& Tyson 2009),
CL J0152-1357 $z=0.84$ (Jee et al. 2005a), MS 1054-0321 $z=0.83$ (Jee et al. 2005b), and RX J0848+4453 $z=1.27$ (Jee et al. 2006).
We refer readers to individual papers for detailed information for each cluster. 

In order to avoid any potential
scatter introduced by variations in the mass determination method, we re-calculated the NFW parameters for these additional clusters
using the same mass-concentration relation employed in the current paper.

\subsection{Dynamical Velocity Dispersion}
Table 2 lists the dynamical velocity dispersions and the predicted values from the lensing analysis for the combined sample.
There are a total of 23 clusters whose velocity dispersions are available either in
the literature or through our collaborations (Meyers et al.  2011; Gilbank et al. in prep.). 

\begin{figure}
\plotone{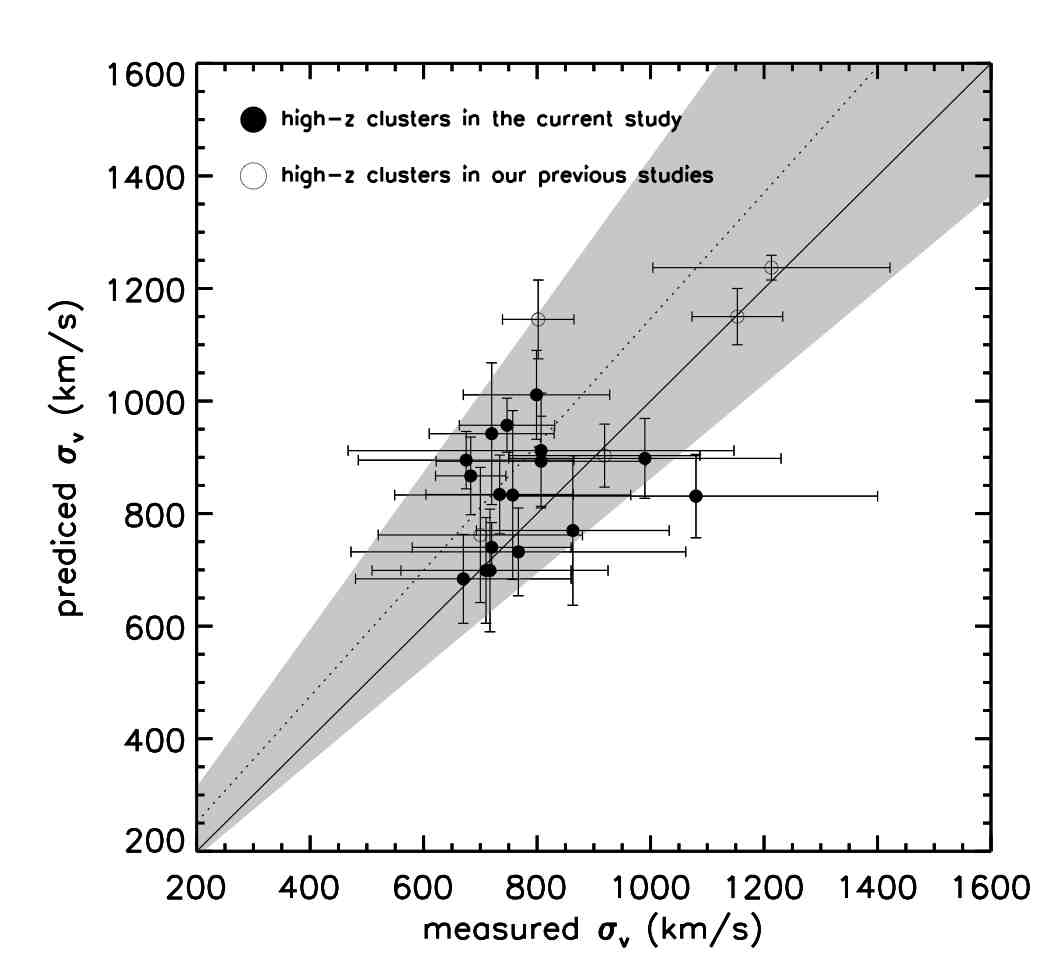}
\caption{Dynamical velocity dispersion versus lensing prediction. We use filled circles to represent the results
for the clusters studied here.
Open circles show the results for the high-$z$
clusters in our previous publications. The solid line represents the line of equality  The linear fit to the result is
shown with the dotted line while the 1-$\sigma$ range of the slope is depicted by the grey region.
\label{fig_velocity_dispersion}}
\end{figure}

Figure~\ref{fig_velocity_dispersion} shows the comparison of the lensing prediction from the SIS fit with the dynamical measurement for these
23 clusters. The filled and open circles represent the clusters analyzed in the current paper and in our previous studies, respectively.
The dotted line is a fit to the data, 
whose slope $\alpha=1.12\pm0.31$ and intercept $b=28\pm260$~$\mbox{km}~\mbox{s}^{-1}$ are consistent with the line of equality (solid); the shadow represents the 1-$\sigma$ range of the slope.
Individual data points have rather large scatters around these lines mostly due to their
large statistical (e.g., small number of known spectroscopic members and source galaxies) and
systematic (e.g., cluster mass profile being different from SIS) uncertainties.

\subsection{X-ray Properties}

Scaling relations between X-ray properties and lensing masses not only provide invaluable
insight into the physical mechanism of cluster formation, but also help us to calibrate
mass estimates based on X-ray data.  
For a virialized cluster, we expect the mass to scale with the gas temperature via the power law relation
$M\propto T^{3/2}$ (Kaiser 1986). In this paper, we use the following specific form to fit the data:

\begin{equation}
E(z)M_{\Delta} = M_5 \left ( \frac{T}{5~ \mbox{keV}} \right )^{\alpha} \label{eqn_scaling},
\end{equation}
\noindent
where $E(z)$ is the redshift-dependent Hubble parameter
\begin{equation}
E(z)=\frac{H(z)}{H_0} = \sqrt{ \Omega_M (1+z)^3 + \Omega_{\Lambda}}.
\end{equation}
 
\begin{figure}
\plotone{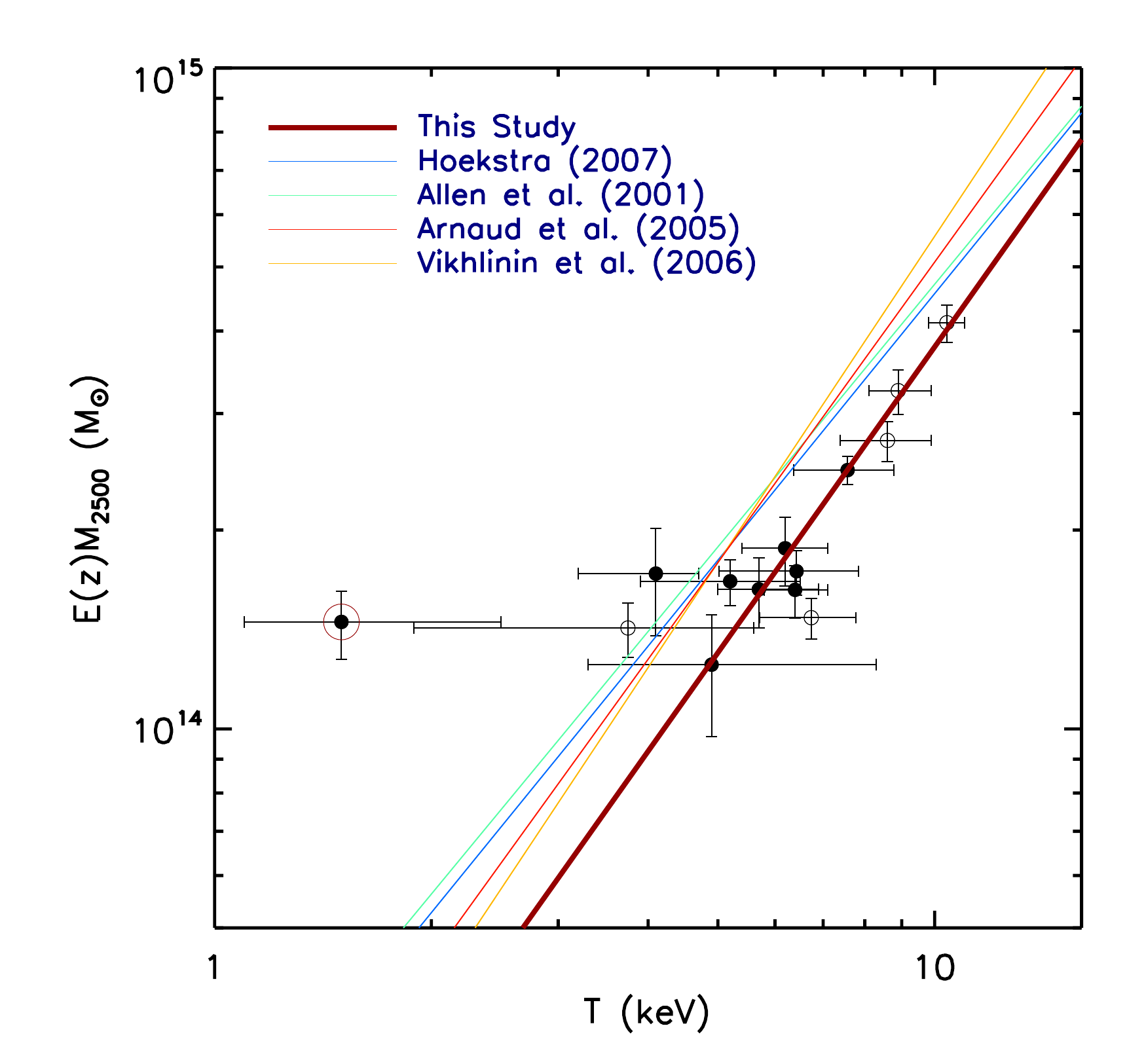}
\caption{X-ray temperature versus lensing mass. As in Figure~\ref{fig_velocity_dispersion}, filled and open circles
represent the current clusters studied here and the five clusters in our previous publications. The slope of the
power law $\alpha=1.54\pm0.23$ ($M\propto T^{\alpha}$) is consistent with the theoretical prediction $3/2$ and also previous results obtained from low redshift samples. Note that the Hoekstra (2007) relation shown here is the revised result after we applied the 10\% reduction in mass as explained in Mahdavi et al. (2008). The greatest outlier from the $M-T_X$ relation is RCS 0439-2904 (red circle), which perhaps is
a line-of-sight superposition of multiple components (Cain et al.  2008) boosting the weak-lensing mass measurement without significantly affecting temperature measurement. 
\label{fig_mass_temperature}}
\end{figure}

In Equation~\ref{eqn_scaling}, the mass $M_{\Delta}$ is usually defined as the total mass within the radius,
at which the mean density becomes $\Delta$ times the critical density. We quote the results
for $M_{2500}$ for an easy comparison with previous studies.
In our combined sample, 14 clusters have published X-ray temperatures (Table 2) spanning the $1.7\mbox{~keV}< T < 10.4 \mbox{~keV}$ range
with reasonable constraints on their uncertainties.
For these clusters, we present the mass-temperature relation in Figure~\ref{fig_mass_temperature}.
The thick red line shows the best-fit power-law relation with $\alpha=1.54\pm0.23$ and $M_5=(9.13\pm0.85)\times10^{13}~h^{-1}~M_{\sun}$.
The cluster RCS~0439-2904 (marked with red circle) is
a significant outlier from this relation and the fit shown here is obtained without including this cluster.
The exclusion of this cluster in our estimation of the M-T relation
is justified because there is a strong indication that RCS 0439-2904 might be a line-of-sight superposition
of multiple clusters (\textsection\ref{section_individual}). However, we stress that adding RCS 0439-2904 does not change our results
(reducing the slope to $1.48\pm0.23$).

Allen et al. (2001) studied six relatively relaxed clusters at $z\lesssim0.46$ observed with $Chandra$ and obtained
the mass-temperature relation $\alpha=1.51\pm0.27$ and $M_5=(1.32\pm0.13)\times10^{14}~h^{-1}~M_{\sun}$.
Arnaud et al. (2005) used six nearby  ($z\lesssim0.15$) $T>3.5$ keV clusters observed with XMM-Newton, and measured the
slope $\alpha=1.51\pm0.11$ and the normalization $M_5=(1.25\pm0.04)\times10^{14}~h^{-1}~M_{\sun}$, which are
consistent with the results of Allen et al. (2001).
A larger sample from $Chandra$ has been studied by Vikhlinin et al. (2006), who analyzed 13 low-redshift clusters in the
temperature range $0.7-9$ keV. They also agreed that the observed scaling relation is in good accordance with
the theoretical prediction, albeit with slightly higher slope $1.64\pm 0.06$ than previous X-ray results.
Their normalization $M_5=(1.25\pm0.05)\times10^{14}h^{-1}~M_{\sun}$ is in good agreement with
previous X-ray results.

Because X-ray studies derive cluster masses from temperature with hydrostatic equilibrium assumption, these results showing that the correlation is
tight and the power law slope of the $M-T_X$ relation is close to the theoretical prediction 3/2 may not be a direct proof for the cluster self-similarity.
It is critical to investigate if the relation still holds when the masses are given by an independent estimator.

The weak-lensing mass versus X-ray temperature relation is investigated by Hoekstra (2007), who derived
$\alpha = 1.34_{-0.28}^{+0.30}$ and $M_5=(1.4\pm 0.2)\times10^{14}~h^{-1}~M_{\sun}$ from the 17 clusters at  $0.17\lesssim  z \lesssim 0.54$, which
are in good agreements with the values from X-ray studies. Mahdavi et al. (2008) updated the redshift distribution used by Hoekstra (2007), and
found that the new $n(z)$ from the photometric redshift catalog of Canada-France Hawaii Telescope Legacy Survey (CFHTLS)
derived by Ilbert et al. (2006) gives on average $\sim$10\% smaller masses. We show this revised result in 
Figure~\ref{fig_mass_temperature}. We further note that  Hoekstra (2007) used the ASCA temperatures that do not correct for cool cores. The absence of this cool-core
correction can bias the temperatures low.

We verified that the above $M-T_X$ relation is not sensitive to a mass-concentration relation. When we recompute our $M_{2500}$ using the mass-concentration
relation of Bullock et al. (2001), the power-law slope $\alpha$ virtually remains the same and the normalization ($M_5$) decreases only $\sim$3\%.
The X-ray and Hoekstra (2007) results were obtained without any assumption on mass-concentration relation. Nevertheless, we tested if the result of Hoekstra (2007)
changes when the masses are derived from NFW profile fitting with the Duffy et al. (2008) mass-concentration relation as is done in this paper. Again, we found that neither the
normalization nor the slope of the $M-T_X$ relation of Hoekstra (2007) changes because of this mass determination method.

It is easy to see in Figure~\ref{fig_mass_temperature} that there is
a $20-30$\% discrepancy in normalization between our result and previous results.
The discrepancy  may imply that the normalization
decreases with redshift or our weak-lensing masses are biased low. One cause for possible weak-lensing bias is the redshift
estimation bias, similar to the case in Hoekstra (2007). We estimate that the sample variance of UDF may be responsible
for 5-12\% shift in mass depending on the cluster redshift. However, although we cannot exclude this possibility,
this would aggravate the already problematic existence of the most massive clusters in the high-z universe (see \textsection\ref{section_abundance} for details).
Therefore, it seems more plausible that there is some evolution in the normalization of the mass-temperature relation to the highest redshifts.

\begin{figure}
\plotone{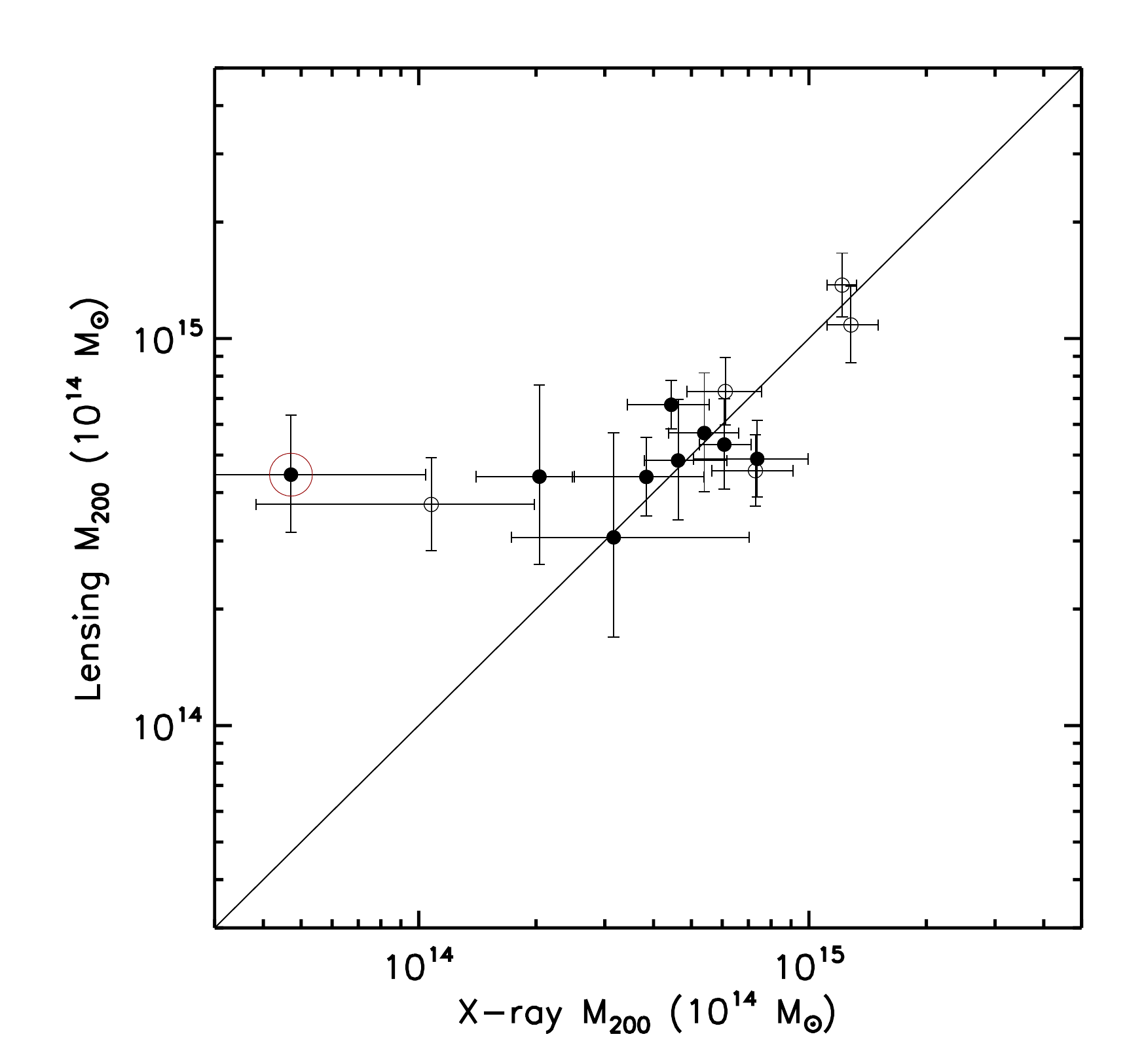}
\caption{Virial mass comparison between X-ray and weak-lensing results. 
Not surprisingly, the trends are similar to that of the mass-temperature relation shown in Figure~\ref{fig_mass_temperature}. 
The cluster RCS0439-2904 (red circle) again is the greatest departure from the line of equality.
Filled and open circles represent the current clusters studied here and the five clusters in our previous publications, respectively.
\label{fig_xmass_vs_lmass}}
\end{figure}

Finally, we compare the masses derived from X-rays and lensing analyses. The cluster masses at $r_{200}$ obtained
from the X-ray results are listed in Table 2. 
We use an isothermal
$\beta$ assumption to derive these values in order to take advantage of the published parameters in the literature.
For the clusters that do not have published measurements of their X-ray
surface brightness profile (because the S/N of the existing data do not constrain the shape of the
X-ray profile well) we assume $\beta=0.7$.
Figure~\ref{fig_xmass_vs_lmass} shows the result for 14 clusters in our combined sample.
The agreement between the lensing and X-ray estimates is good except for RCS0439-2904, which, as mentioned already,
might be a line-of-sight superposition of multiple components (Cain et al. 2008). It is important to remember that
an SIS profile is assumed for the derivation of $M_{200}$ from X-ray data. Therefore, it is possible that this excellent agreement might be
a coincidence to some extent and a systematic difference might appear if masses are evaluated at different radii.
 
\section{ABUNDANCE OF THE MOST MASSIVE CLUSTERS AT HIGH REDSHIFT }\label{section_abundance}

The abundance of the most massive clusters at high redshift is extremely sensitive to cosmological parameters. In the current era when we can estimate the cosmological parameters to a high precision [mainly from the cosmic microwave background (e.g., Komatsu et al. 2011) and Type-Ia supernova studies (e.g., Amanullah et al. 2010)],  discovery of even a single massive 
cluster can challenge the current $\Lambda$CDM model (Mortonson et al. 2011).

In Jee et al. (2009), we presented a weak-lensing analysis of XMM J2235-2557 and found that the mass of the cluster
is surprisingly high. With the cosmological parameters fixed at the current
best-fit values, the theoretical probability of finding such a massive cluster in the survey volume is only $\lesssim$1\%. The exact value for the probability
depends on the adopted mass function and cosmological parameters. Nevertheless,  independent estimates (e.g., Jimenez \& Verde 2009; 
Sartoris et al. 2010; Holz \& Perlmutter 2010) agree that the discovery provides significant tension with the current $\Lambda$CDM cosmology.
In this section, we extend the study of Jee et al. (2009) to the most massive clusters in our sample with a new mass function, 
revised survey areas, and marginalization over the uncertainties of cluster masses and cosmological parameters. 

\subsection{Mass Function}
The mass function is often given in the following form:
\begin{equation}
\frac{dn}{d \ln M} = \frac{\rho_{m,0}}{M} \left | \frac{d \ln \sigma(M,z)} {d \ln M} \right | f,
\end{equation}
\noindent
where $\sigma$ is the rms variation of the density field when smoothed on scale $M$. $\rho_{m,0}$ is the present matter density.
The function $f$ determines the shape of the mass function.

Jenkins et al. (2001) proposed a redshift-independent form $f=f(\sigma)$, which makes the above mass function agree with their
numerical simulation results with high precision. Tinker et al. (2008) improve the agreement by considering the redshift-dependence
of $f=f(\sigma,z)$, which is expressed as
\begin{equation}
f(\sigma,z)=A \left [ \left ( \frac{a}{b} \right ) ^{-a} + 1 \right ]  e^{-c/\sigma^2}. 
\end{equation}
\noindent
where $A=0.186(1+z)^{-0.14}$, $a=1.47(1+z)^{-0.06}$, $b=2.57(1+z)^{-0.011}$, and $c=1.19$ when the mass is defined
as the total mass at the radius, inside of which the mean density is 200 times the mean density of the universe at the cluster redshift.
We refer to it as $M_{200m}$ when necessary to distinguish it from $M_{200}$, which is more commonly
used to refer to the mass at the radius, where the mean density becomes 200 times the $critical$ density.
The deviation of the universality of the mass function by Jenkins et al. (2001) from the numerical simulation results  is 20-50\% (Tinker et al. 2008);
Bhattacharya et al. (2011) argue that the difference between different fitting formulae is ÊprobablyÊ larger than the scatter seen in the largest
simulations.

\subsection{Cluster Survey Area}\label{section_survey_area}

The cluster survey area that one chooses to adopt to evaluate the cosmological significance for discovery of unusually massive
clusters is somewhat controversial. For example, Mortonson et al. (2011) argue that $\sim$300 deg$^2$ should be used
for the search area of XMM2235-2557 at $z=1.4$ instead of the pilot survey area $\sim11$ deg$^2$ (Jee et al. 2009).   
The argument for this $\sim300$ deg$^2$ area is based on the study of Hoyle et al. (2011), who
assumed $\sim$168 deg$^2$ for XCS, $\sim$64 deg$^2$ for XMM-LSS, $\sim$11 deg$^2$ for XDCP, $\sim$17 deg$^2$ for WARPS, and
$\sim$33 deg$^2$ for RDCS. However, the different XMM-Newton-based surveys overlap significantly. In addition,
because the flux limit that enables a secure cluster detection at $z>1$ is $\sim10^{-14} \mbox{erg~cm}^{-2}\mbox{s}^{-1}$, the effective search area must be
reduced significantly. For example, the XDCS survey has continued since the end of the 11 deg$^2$ pilot survey and now reaches
$\sim50$ deg$^2$ (R. Fassbender in prep.) for a flux limit $\sim10^{-14} \mbox{erg~cm}^{-2}\mbox{s}^{-1}$.
The RDCS survey covers a geometrical area of 50 deg$^2$ only at relatively
high fluxes. However the area suitable for $z>1$ cluster detection is
$\sim$5 deg$^2$, corresponding to fluxes below $\sim3\times10^{-14}
\mbox{erg~cm}^{-1}\mbox{s}^{-1}$  (Rosati et al. 1998).
Considering the argument listed above,  the combined survey area for the X-ray selected high-$z$ clusters should be $\sim$100 deg$^2$.

\subsection{Abundance Estimation and Eddington Bias}

In the full sky survey, the number of clusters with mass and redshift greater than $M_{min}$ and $z_{min}$, respectively is given by
\begin{equation}
N(M,z) = \int_{z_{min}}^{z_{max}}  \frac{dV(z)}{dz} dz \int_{M_{min}}^{M_{max}}  \frac{dn}{dM}  dM \label{eqn_abundance}
\end{equation}
\noindent
where $dV/dz$ is the volume element and $dn/dM$ is the mass function. For massive high-redshift clusters, the result
is sensitive to the mass function $dn/dM$ near $z_{min}$ and $M_{min}$.

We compute the probability of finding each cluster by marginalizing over the measurement errors, assuming
a log-normal distribution for $P(M)$. This is different from the 
approach of Jee et al. (2009), where we
simply used a threshold mass $M_{thr}$ to evaluate the integral (Eqn~\ref{eqn_abundance}).
Nevertheless, we emphasize that the new method yields values
very close to those obtained from the previous method when the mass uncertainty is relatively small ($\lesssim 20$ \%).
For example, marginalizing over the measurement error gives an abundance of $\sim2\times10^{-3}$
for XMM J2235-2557 in the original 11 deg$^2$ XDCP survey, which is in good agreement with the previous value $\sim5\times10^{-3}$.
As we use a fixed value of $z_{max}$, the resulting abundance is slightly conservative because
in principle the maximum redshift  should decrease at the low end of $P(M)$; remember that at the
high end of $P(M)$, the steep mass function cancels the effect of increased $z_{max}$. 
For the conversion from expected number to probability we assume Poissonian distribution. That is, if the expected
number is $N$ within a particular survey, the discovery probability is simply $P=1-e^{-N}$. 

Mortonson et al. (2011) noted that an Eddington bias (Eddington 1913) is an important factor in the discussion of the cosmological significance using
most massive clusters. Eddington bias is the combined effect of a steep mass function and measurement uncertainty. 
For example, consider the weak-lensing mass estimate of XMM2235  $M_{200}=(7.3\pm1.3)\times10^{14} M_{\sun}$. Because the slope of the mass function near the cluster mass is steep, there should be more $\lesssim7.3\times10^{14} M_{\sun}$ objects
scattering up than $\gtrsim7.3\times10^{14} M_{\sun}$ objects scattering down. As a matter of course, this does not mean that XMM2235's mass should be
quoted with a lower mass (for individual clusters the scatter is symmetric). However, when we discuss the existence of the cluster
in probabilistic terms, this additional chance of bias should be considered from a Bayesian point of view.

We adopt the convenient Mortonson et al. (2011) prescription, which is to replace the observed mass with the following reduced mass $M^{\prime}$ in the estimation of the abundance: $M^{\prime}=\exp(1/2 \gamma \sigma_{\ln M} )M$,  where $\gamma$
is the local power-law slope ($dn/d\ln M \sim M^{\gamma}$),  and $\sigma_{\ln M}$ is the 1-$\sigma$ uncertainty
of $\ln M$ (log-normal distribution is assumed for mass errors).

\subsection{Probability of Discovery for Most Massive Clusters}

\begin{figure}
\plotone{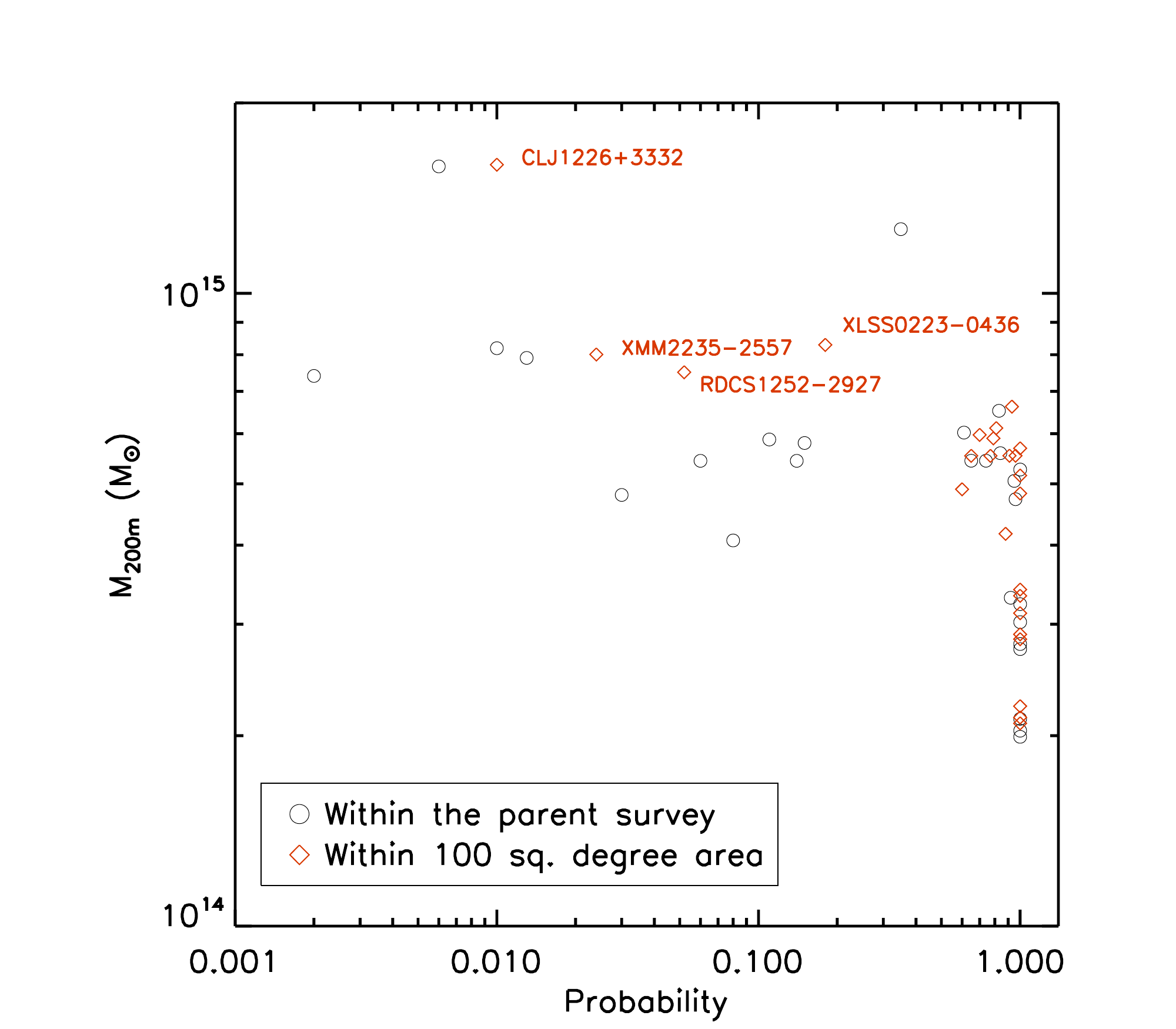}
\caption{Discovery probability of clusters. After applying Eddington bias to the lensing masses, we marginalize over
the uncertainties of the WMAP7 cosmological parameters in order to estimate the discovery probability.
MS 1054-0321 and CL J0152-1357 were discovered in the Einstein Medium Sensitivity Survey (EMSS), which covered a total of
778 deg$^2$ (Gioia et al. 1990). We do not recompute the probability for these two clusters with 100 deg$^2$.
\label{fig_discovery_probability}}
\end{figure}

In Figure~\ref{fig_discovery_probability} and Table 3 we show the discovery probability of all clusters (i.e., similarly massive clusters) studied in this paper. 
We compute the probabilities after marginalizing over the Wilkinson Microwave Anisotropy Probe 7-year (WMAP7) Monte Carlo Markov Chain (MCMC) data\footnote{available at http://lambda.gsfc.nasa.gov/product/map/current/params/lcdm\_sz\_lens\_wmap7.cfm.}.
The probabilities estimated within the parent survey may lead to a
somewhat aggressive interpretation for some clusters. For example, most of the RDCS clusters have
significantly low probabilities mainly due to the small ($\sim$5 deg$^2$) survey area. 
However, most  non-X-ray selected clusters (ISCS and RCS) have discovery probability equal or close to unity
even within their parent surveys.

As mentioned in \textsection\ref{section_survey_area}, we believe that the survey area 100 deg$^2$ is still a conservative choice for most of the clusters shown here (the exceptions are MS1054-0321 and CL J0152-1357, which were discovered in the EMSS $\sim735$ deg$^2$ survey).  With this survey area, there remain four clusters whose discoveries are still unusual given our current understanding of cosmological parameters (the names of these clusters are
shown in Figure~\ref{fig_discovery_probability}). The cluster with the lowest probability ($\sim$1\%) is the $z\sim0.9$ cluster CLJ1226+3332, whose weak-lensing mass and X-ray temperature are $M_{200m}\sim1.6\times10^{15}M_{\sun}$ and $T_{X}\sim10$ keV, respectively. Because the cluster was discovered in a $72$ deg$^2$ survey (WARPS), using 100 deg$^2$ does not increase the probability significantly. In addition, XMM-Newton archival search programs look for regions where no known X-ray luminous clusters exist, and thus the search volume in the XMM-Newton archival survey excludes the $z\lesssim1$ volume by observer's selection. 
This makes our probability estimation for CLJ1226+3332 even more conservative.
The second lowest probability is found with XMM2235-2557 at $z=1.4$. The probability $\sim$3\% is much higher (after the area and Eddington bias correction) than our previous value $\sim$0.5\% within the 11 deg$^2$ pilot XDCP survey, but still gives $\sim2 \sigma$ tension with the predicted number of clusters from $\Lambda$CDM.
RDCS1252-2927 at $z=1.24$ is ranked as third in rarity, giving $P\sim$6\%, which is followed by XLSS0223-0436 at $z=1.22$ ($P\sim$18\%).

Combining discovery probabilities of all our clusters is somewhat tricky. Because of Poissonian fluctuation, the discovery probability
is always lower than the estimated abundance (i.e., $P=1 - e^{-N}$). For example, if the expected number of a given cluster
is $N=1$ within 100 deg$^2$ survey, the probability is $\sim0.63$. Obviously, simply multiplying probabilities of all clusters potentially leads to
an unreasonably low final value if there are many slightly-less-than-one-probability clusters. Nevertheless, if we include only rare clusters ($P<<1$),
multiplying individual probabilities provides a useful measure on how much the existence of these multiple clusters give
rise to tension with the current cosmological parameters. 

One set of cosmological parameters that increase the predicted abundance of one massive cluster can boost another similarly massive cluster's abundance.
Therefore, multiplication of probabilities must be done separately for each chain, and the final probability should be obtained after
properly taking into account each chain's weight.  We find that the final combined probability obtained in this way is $\sim$0.03\%. 

\subsection{Exclusion Curve Test}

Mortonson et al. (2011) presented fitting formulae, which provide an exclusion mass as a function of redshift
for given sample and parameter variance confidence limits (CL). 
Even a single cluster equal to or above the exclusion mass
rules out both $\Lambda$CDM and quintessence.

\begin{figure}
\plotone{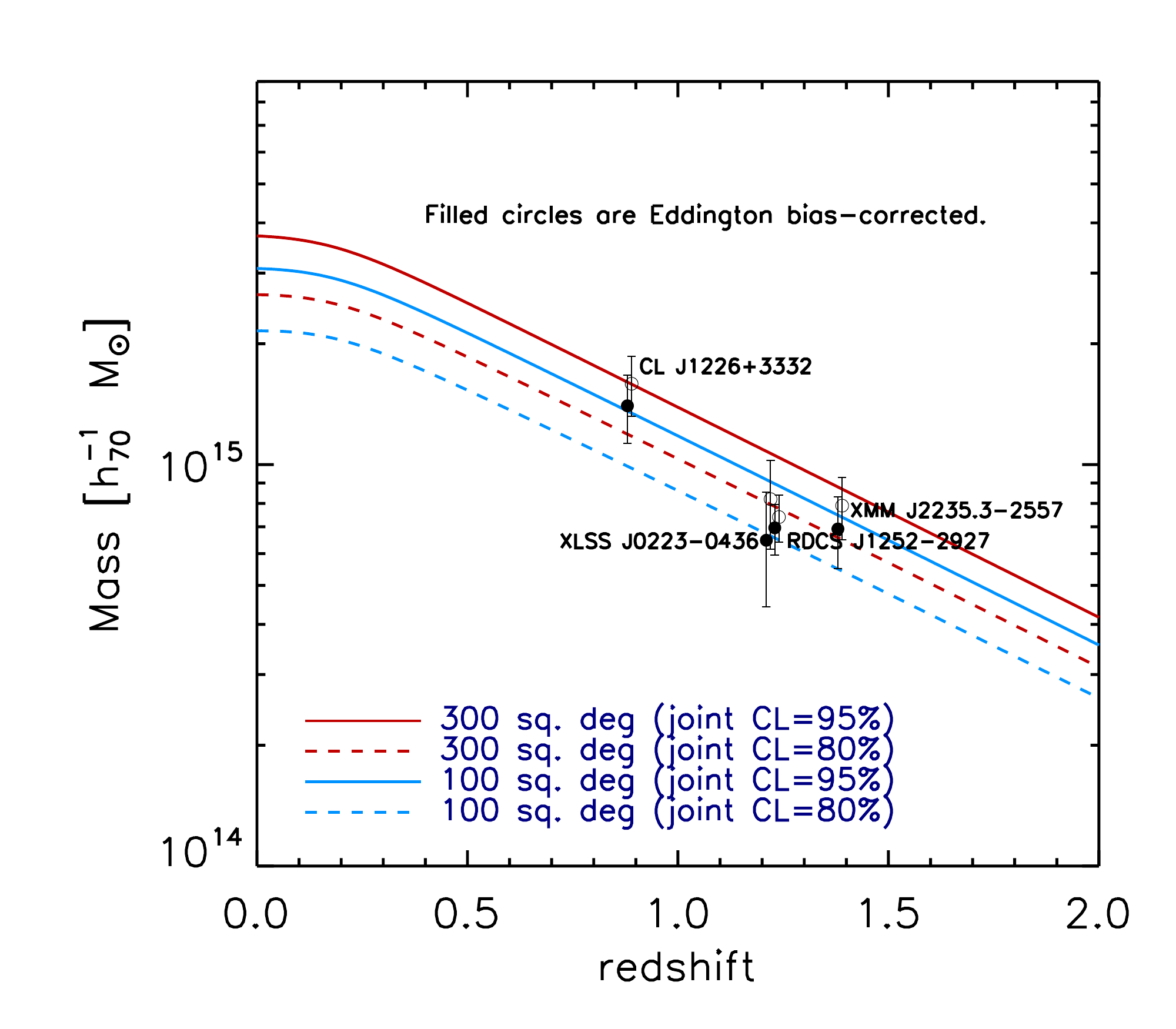}
\caption{Exclusion curves. Any cluster's mass above the given curve excludes the $\Lambda$CDM
with the joint CL specified (Mortonson et al.  2011). For example, the central value (Eddington bias-corrected) of CL J1226+3332
rules out $\gtrsim$95\% of currently allowed $\Lambda$CDM models at the $\gtrsim95$\% confidence level for the
100 deg$^2$ area.
\label{fig_exclusion}}
\end{figure}

In Figure~\ref{fig_exclusion}, we compare our clusters to the exclusion curves of Mortonson et al. (2011). Although we believe that
the proper choice for the sky area is 100 deg$^2$ or less, we also plot the result for the 300 deg$^2$ assumption, which was
adopted in Mortonson et al. (2011).  For simplicity, we only consider the case where the sample variance CL is equal to
the parameter variance CL, which is referred to as joint CL in Mortonson et al. (2011).

The central mass of the cluster CL J1226+3332 after the Eddington bias correction lies above the joint CL 95\% exclusion curve
when we adopt 100 deg$^2$ for the survey area. All the masses of the four clusters shown here are approximately
on or above the joint CL 80\% exclusion curves. When we consider 300 deg$^2$ for the search area instead, we note
that the two clusters CLJ1226+3332 and XMM J2235-2557 are above the 80\% exclusion curves. We clarify here that
Mortonson et al. (2011) used the mass of XMM J2235-2557 reported by Stott et al. (2011), whose value is
similar to our weak-lensing mass, but has significantly larger uncertainty ($\sim40$\%); Stott et al. (2011) derived
the cluster mass from their X-ray scaling relation.
This large uncertainty leads to $\sim40$\% Eddington bias, and Mortonson et al. (2011) concluded that
the resulting mass of the cluster provides only a negligible tension with $\Lambda$CDM when they assume 300 deg$^2$
for the sky area. On the other hand, our mass uncertainty is $\sim18$\%, and the corresponding
Eddington bias is $\sim12$\%, which makes the corrected mass still close to the $\sim$85\% exclusion curve for Mortonson et al. (2011)'s
300 deg$^2$ assumed area.

The above exclusion curve test provides weaker constraints when one is using more than a single cluster to examine their cosmological significance because
the tool adopts the lowest redshift for the sample in the estimation of the cluster abundance (Mortonson et al. 2011).  Therefore, we do not attempt to estimate
the tension by combining the four most massive clusters.

\subsection{Interpretation of Overabundance of Massive Clusters at High Redshift}

Assuming the current well-accepted $\Lambda$CDM cosmology, we have demonstrated above that
the discovery of the most massive clusters in our sample are rare events. Among the most popular
interpretation would be that the primordial density fluctuation is non-Gaussian.  For example, Jimenez \& Verde (2009)
claim that the local $f_{NL}$ should be in the range 150-260 to accommodate XMMU2235-2557 within the pilot 11 deg$^2$
survey.  We suspect that their estimate of $f_{NL}$ would decrease somewhat when both the Eddington bias and
revised area $50-100$ deg$^2$ are taken into account.
A tighter constraint is possible if one utilizes the weak-lensing masses of the four most massive clusters studied here.
A similar study is performed by Hoyle et al. (2011) from the investigation of the 14 high-redshift cluster masses reported in the literature,
giving a significantly large $f_{NL}>375$ (after marginalization over WMAP5 parameters).  
 
One crucial test that however should accompany these non-Gaussianity studies is the investigation of the  mass function
at the high mass end. The existing limited-volume simulations do not constrain the number density of extremely massive clusters
accurately, and the commonly used fitting functions are simply extrapolated results in this regime.
Another potentially important contribution from numerical simulation is the predicted mass function specific for a given survey,
which takes into account the various aspects of selection limits and projection effects.

The projection effect is always a concern in the cosmological interpretation of extremely massive clusters. Certainly, a
superposition of two moderately massive clusters or a long filament viewed along the line of sight can be identified with an extreme object.
The former case is easily detected by a redshift histogram
if the field is spectroscopically surveyed. Initially, RDCS1252-2927 at $z=1.24$ was such a candidate because the redshift
catalog showed three possible foreground groups at at $z=0.47$, 0.68, and 0.74 (Lombardi et al. 2005). However, the spatial
distributions of these groups are not compact, and they appear to be just loose galaxy groups commonly found in any random field.
The latter case (a filament viewed along the line of sight)  is very difficult to prove (or disprove) directly. Nevertheless, there are a few circumstantial
lines of evidence, which help us to infer if a system possesses extreme triaxiality.
Comparing X-ray temperature with weak-lensing mass is useful because a severe triaxiality will give rise to
much higher weak-lensing mass than what the temperature predicts.  In addition, a line-of-sight elongation
inflates velocity dispersion, and the result will be significantly larger than what the lensing predicts.
XLSS0223-0436 is potentially such an object because our lensing mass $M_{200}=7.4_{-1.8}^{+2.5}\times10^{14}~M_{\sun}$  is
much larger than what the X-ray temperature $\sim3.8$~keV (Bremer et al. 2006) suggests.  However, the X-ray photon statistics
are poor, and Bremer et al. (2006) could not determine an upper limit of the X-ray temperature. Deeper Chandra observations would be necessary
to robustly examine this large (a factor of 2-3) discrepancy. We note that the existing data for CLJ1226+3332 and XMM J2235-2557, the two clusters with the largest
tension with $\Lambda$CDM in our sample, do not suggest the possibility that weak-lensing masses are significantly boosted by any projection effect.

\section{SUMMARY AND CONCLUSIONS}  \label{section_conclusion}

We have presented a weak-lensing analysis of 22 high-redshift ($0.91<z<1.46$) clusters using HST/ACS images. 
Most clusters in the sample show a significant lensing signal and are well detected in our two-dimensional mass reconstruction.
We complement our sample with five high-redshift clusters that were studied in the past in order to compare
lensing results with X-ray and dynamical measurements over a large range of masses, and to discuss the most massive
clusters in this combined sample in the context of cosmology.
Our results are summarized as follows.

\begin{enumerate}
\item The dynamical velocity dispersions are consistent with the values estimated by lensing analysis. 
\item The power law slope of the lensing mass-temperature relation is in good agreement with the theoretical prediction, which is also
supported by previous investigations carried out with X-ray data alone or with both X-ray and lensing data.
\item We see evidence of an evolving normalization constant in the $M-T_X$ relation.
\item The masses derived from X-ray and lensing measurements are in good agreement.
\item The existence of the most massive clusters in our sample gives rise to tension with the current $\Lambda$CDM standard structure
formation paradigm. 
\end{enumerate}

The first conclusion is consistent with the findings of Hoekstra (2007) who studied low redshift clusters. Our result demonstrates that the relation
can be extended to this high-redshift regime, albeit with slightly increased scatter. This larger scatter could be intrinsic
for high redshift clusters, which are likely to possess more merging substructures than
low redshift systems. However, one must remember that most clusters in our sample do not have sufficient number of
spectroscopic members yet to give reliable estimates. In addition, the depth of the ACS images is also
less than optimal in many cases, which increases uncertainties for the lensing values. Our assumption of SIS
in the derivation of the velocity dispersion may introduce some bias in the relation simply because the
real clusters are not singular isothermal. Nevertheless, the collective
behavior of the relation being consistent with the equality is still noteworthy, and hints at the possibility that the
dynamical measurements from a large number of cluster members
may still serve as useful proxies for masses in this redshift regime.

The mass-temperature relation presented here is tight, particularly for hot clusters.
The current result indicates that the power-law slope of the mass-temperature relation does not
evolve since $z\sim1$. However, our normalization value is $20-30$\% lower than previous results, suggesting that
the X-ray temperatures of the clusters at $z\sim1$ may be systematically higher than at low redshifts. The shift
can imply that non-thermal sources of heating (e.g., merging, AGN activity, etc.) might be more prevalent at higher redshift, or simply
the difference in gas fraction (because of different cluster formation epoch) might be significant.
Alternatively, one can also consider  bias in our lensing analysis. However, it is very unlikely that 
our calibration error, even at a maximum, can induce $20-30$\% systematic errors. Our shear calibration error verified by
numerical simulations is at most at the $\sim1$\% level. The redshift estimation bias (i.e., the magnitude versus redshift relation in the UDF
does not represent the mean relation of the universe) is at the $5-12$\% level for each cluster, giving the highest error for the most distant object.
Most importantly, if the cluster masses in our sample should be somehow increased by $20-30$\%, this would not only degrade the
dynamic versus lensing velocity dispersion relation substantially, but also make the existence of the most massive clusters much more problematic with
the current cosmological parameters.

The fourth conclusion is somewhat foreseen by the tight mass-temperature relation. However, this direct comparison of mass between
X-ray and lensing methods provides another way to verify our normalization in the $M-T$ relation. Again, the clusters with the lowest 
temperatures are the greatest outliers from the equality. On the other hand, the other relatively massive clusters 
give consistent results between X-ray and lensing measurements. When we repeat the analysis by re-calculating the cluster mass at
smaller radii (e.g., $M_{500}$ instead of $M_{200}$), the relation virtually remains the same. This indicates that the difference
in the assumption of the cluster mass profile (i.e, isothermal $\beta$ versus NFW) is not a significant source of bias between the two
measurements. 

Finally, the last conclusion is the most striking finding of the current investigation. The above studies on the relations between
lensing mass and other properties do not support the possibility that our lensing mass determination is significantly biased high.
A significantly high degree of non-Gaussianity or high value of $\sigma_8$  may explain the existence of these massive clusters
at high redshift. However, we caution that our current theoretical knowledge of the mass function at high end is limited by small number
statistics.

Foley et al. (2011) reported the discovery of SPT-CL J2106-5844, which is claimed to be the most massive cluster known at $z>1$
based on the X-ray temperature and SZ measurement. The authors estimate that there is only a 7\% chance of finding such a
massive cluster within the 2,500 deg$^2$ survey area. The mass of the cluster still needs to be verified by other methods (e.g., weak lensing analysis).
However, if the cluster is indeed massive as reported, the existence of SPT-CL J2106-5844 also favors our claim that the most
massive clusters at high redshift provide non-negligible tension with our current understanding of structure formation.

M. J. Jee acknowledges support for the current research from the TABASGO foundation presented in the form of
the Large Synoptic Survey Telescope Cosmology Fellowship.
Financial support for this work was in part provided by NASA through program
GO-10496 from the Space Telescope Science Institute, which is operated
by AURA, Inc., under NASA contract NAS 5-26555.  This work was also
supported in part by the Director, Office of Science, Office of High
Energy and Nuclear Physics, of the U.S. Department of Energy under
Contract No. AC02-05CH11231, as well as a JSPS core-to-core program
``International Research Network for Dark Energy'' and by JSPS
research grant 20040003.  H. Hoekstra acknowledges support from the Netherlands Organisation for Scientific
Research (NWO) through a VIDI grant. H. Hoekstra is also supported by a Marie Curie International Reintegration Grant.
P. Rosati acknowledges partial support by the DFG cluster of
excellence Origin and Structure of the Universe.
Support for M. Brodwin was provided by the
W. M. Keck Foundation.  The work of S~.A~.Stanford was performed under the
auspices of the U.S. Department of Energy by Lawrence Livermore
National Laboratory in part under Contract W-7405-Eng-48 and in part
under Contract DE-AC52-07NA27344.  The work of P. Eisenhardt was
carried out at the Jet Propulsion Laboratory, California Institute of
Technology, under a contract with NASA.

\clearpage

\clearpage

\clearpage

\clearpage

\begin{deluxetable}{lccccccc}
\tabletypesize{\scriptsize}
\tablecaption{Details of HST observations}
\tablenum{1}

\tablehead{\colhead{Cluster name} & \colhead{Redshift} & \colhead{RA} &
\colhead{DEC} & \colhead{F775W exp.} & \colhead{F850LP exp.} & \colhead{$n_{bg}$} & \colhead{Prop. ID} \\
\colhead{} & \colhead{} & \colhead{(hh mm ss)} & \colhead{($~~\degr~~~\arcmin~~~\arcsec$)} &
\colhead{(s)} & \colhead{(s)} & \colhead{(per sq. arcmin)} & \colhead{} \\  }
\tablewidth{0pt}
\startdata
XMMXCS J2215-1738 & 1.46 & 22 15 59 &  -17 37 58 & 3320 & 16935 & 85  & 10496 \\
XMMU J2205-0159   & 1.12 & 22 05 50 &  -01 59 30 & 4535 & 11380 & 77  & 10496 \\
XMMU J1229+0151   & 0.98 & 12 29 29 &  +01 51 22 & 4110 & 10940 & 99  & 10496 \\
WARPS J1415+3612  & 1.03 & 14 15 11 & +36 12 04 & 2425 & 9920 & 78  & 10496 \\
ISCS J1432+3332 & 1.11 & 14 32 29 & +33 32 48 & 4005 & 12440 & 115 & 10496 \\  
ISCS J1429+3437 & 1.26 & 14 29 18 & +34 37 25 & 2670 & 15600 & 81  & 10496 \\
ISCS J1434+3427 & 1.24 & 14 34 28 & +34 26 22 & 2685 & 13320 & 84  & 10496 \\
ISCS J1432+3436 & 1.35 & 14 32 38 & +34 36 49 & 2235 & 8940  & 81  & 10496 \\
ISCS J1434+3519 & 1.37 & 14 34 46 & +35 19 45 & 1920 & 11260 & 70  & 10496 \\
ISCS J1438+3414 & 1.41 & 14 38 09 & +34 14 19 & 2155 & 10620 & 76  & 10496 \\
RCS 0220-0333     & 1.03 & 02 20 55 & -03 33 10 & 2955 & 14420 & 96  & 10496 \\
RCS 0221-0321     & 1.02 & 02 21 41 & -03 21 47 & 2015 & 13360 & 79  & 10496 \\
RCS 0337-2844     & 1.1  & 03 37 50 & -28 44 28 & 1560 & 12885 & 67  & 10496 \\
RCS 0439-2904     & 0.95 & 04 39 38 & -29 04 55 & 2075 & 15530 & 81  & 10496 \\
RCS 2156-0448     & 1.07 & 21 56 42 & -04 48 04 & 2060 & 5440  & 50  & 10496 \\
RCS 1511+0903     & 0.97 & 15 11 03 & +09 03 15 & 2075 & 7120  & 65  & 10496 \\
RCS 2345-3632     & 1.04 & 23 45 27 & -36 32 50 & 4450 & 9680  & 103 & 10496 \\
RCS 2319+0038     & 0.91 & 23 19 53 & +00 38 13 & 2400 & 6800  & 61  & 10496 \\
XLSS J0223-0436   & 1.22 & 02 23 03 & -04 36 18 & 3380 & 14020 & 97  & 10496 \\
RDCS J0849+4452   & 1.26 & 08 48 56 & +44 52 00 &15630 & 34840 & 180 & 9919 and 10496 \\
RDCS J0910+5422   & 1.11 & 09 10 44 & +54 22 08 & 9825 & 25380 & 148 & 9919 and 10496 \\
RDCS J1252-2927   & 1.23 & 12 52 54 & -29 27 17 &29945 & 57070 & 165 & 9290 and 10496 \\
\enddata
\end{deluxetable}

\begin{deluxetable}{lcccccccc}
\tabletypesize{\scriptsize}
\tablecaption{Cluster Properties}
\tablenum{2}
\tablehead{\colhead{Cluster name} & \colhead{$z$} & \colhead{$\sigma_{vel}$} & \colhead{$\sigma_{vel}^{len}$} &
\colhead{$T$} & \colhead{M$^{X}_{200}$} & \colhead{$c_{200}$} & \colhead{$r_{200}$} & \colhead{M$^{L}_{200}$}  \\
\colhead{} & \colhead{} & \colhead{(km/s)} &  \colhead{(km/s)} &\colhead{(keV)} &
\colhead{($10^{14}~M_{\sun}$)} & \colhead{} & \colhead{(Mpc)} & \colhead{($10^{14}~M_{\sun}$)} \\  }
\tablewidth{0pt}
\startdata
XMMXCS J2215-1738   & 1.46 & $720\pm110$\tablenotemark{a} & $942_{-126}^{+111}$  & $4.1_{-0.9}^{+0.6}$\tablenotemark{1}  & $2.0_{-0.6}^{+0.5}$ & $2.45\pm0.11$&  $0.90_{-0.14}^{+0.17}$  & $4.3_{-1.7}^{+3.0}$  \\
XMMU J2205-0159     & 1.12 & $         $ & $679_{-79}^{+71}  $  & $                 $  & $                 $        & $2.72\pm0.10$&  $0.90_{-0.12}^{+0.14}$  & $3.0_{-1.0}^{+1.6}$  \\
XMMU J1229+0151     & 0.98 & $683\pm62$\tablenotemark{b} & $867_{-69}^{+64}  $  & $6.4_{-0.6}^{+0.7}$\tablenotemark{2}  & $5.7_{-0.8}^{+1.0} $ &$2.67\pm0.06$&  $1.12_{-0.10}^{+0.11}$  & $5.3_{-1.2}^{+1.7}$  \\
WARPS J1415+3612    & 1.03 & $807\pm185$\tablenotemark{c} & $893_{-80}^{+73}$    & $5.7_{-0.7}^{+1.2}$\tablenotemark{3}  & $4.6_{-0.8}^{+1.5} $&$2.66\pm0.08$&  $1.09_{-0.12}^{+0.14}$  & $4.7_{-1.4}^{+2.0}$  \\
ISCS J1432+3332     & 1.11 & $734\pm115$\tablenotemark{d} & $834_{-70}^{+64}$    & $                 $  & $                 $      &$2.61\pm0.06$&  $1.06_{-0.09}^{+0.11}$  & $4.9_{-1.2}^{+1.6}$  \\  
ISCS J1429+3437     & 1.26 & $767\pm295$\tablenotemark{e} & $732_{-78}^{+70}$    & $                 $  & $                 $    &$2.51\pm0.08$&  $1.04_{-0.12}^{+0.13}$  & $5.4_{-1.6}^{+2.4}$  \\
ISCS J1434+3427     & 1.24 & $863\pm170$\tablenotemark{e} & $770_{-133}^{+113}$  & $                 $  & $                 $    &$2.68\pm0.14$&  $0.82_{-0.14}^{+0.19}$  & $2.5_{-1.1}^{+2.2}$  \\
ISCS J1432+3436     & 1.35 & $807\pm340$\tablenotemark{e} & $912_{-102}^{+92} $  & $                 $  & $                 $   &$2.46\pm0.08$&  $1.00_{-0.12}^{+0.14}$  & $5.3_{-1.7}^{+2.6}$  \\
ISCS J1434+3519     & 1.37 &                   & $812_{-166}^{+137}$  & $                 $  & $                 $   &$2.59\pm0.15$&  $0.81_{-0.16}^{+0.21}$  & $2.8_{-1.4}^{+2.9}$  \\
ISCS J1438+3414     & 1.41 & $757_{-208}^{+247}$\tablenotemark{f} & $833_{-150}^{+127}$  & $4.9_{-1.6}^{+3.4}$\tablenotemark{4}  & $3.2_{-1.4}^{+3.9} $    &$2.55\pm0.13$&  $0.82_{-0.14}^{+0.18}$  & $3.1_{-1.4}^{+2.6}$  \\
RCS 0220-0333       & 1.03 &      	   	 & $881_{-74}^{+68}$    & $                 $  & $                 $         &$2.66\pm0.07$&  $1.09_{-0.11}^{+0.12}$  & $4.8_{-1.3}^{+1.8}$  \\
RCS 0221-0321       & 1.02 & $ 710\pm150\tablenotemark{g}     $  & $699_{-94}^{+83}$    & $                 $  & $                 $     &$2.90\pm0.13$&  $0.80_{-0.13}^{+0.16}$  & $1.8_{-0.7}^{+1.3}$  \\
RCS 0337-2844       & 1.1  & $         $ & $863_{-113}^{+100}$  & $                 $  & $                 $         &$2.61\pm0.10$&  $1.08_{-0.15}^{+0.17}$  & $4.9_{-1.7}^{+2.8}$  \\
RCS 0439-2904       & 0.95 & $1080\pm320$\tablenotemark{h} & $831_{-74}^{+68}$    & $1.5_{-0.4}^{+1.0}$\tablenotemark{5}  & $0.46_{-1.7}^{+6.0}$    &$2.73\pm0.08$&  $1.09_{-0.11}^{+0.13}$  & $4.3_{-1.2}^{+1.7}$  \\
RCS 2156-0448       & 1.07 & $         $ & $691_{-172}^{+137}$  & $                 $  & $                 $         &$2.86\pm0.20$&  $0.78_{-0.18}^{+0.26}$  & $1.8_{-1.0}^{+2.5}$  \\
RCS 1511+0903       & 0.97 & $717\pm208$\tablenotemark{e} & $699_{-109}^{+94}$   & $                 $  & $                 $  &   $2.92\pm0.13$&  $0.82_{-0.14}^{+0.17}$  & $1.9_{-0.8}^{+1.4}$  \\
RCS 2345-3632       & 1.04 & $670\pm190$\tablenotemark{i} & $684_{-79}^{+71}$    & $                 $  & $                 $  &    $2.81\pm0.09$&  $0.87_{-0.10}^{+0.11}$  & $2.4_{-0.7}^{+1.1}$  \\
RCS 2319+0038       & 0.91 & $990\pm240$\tablenotemark{j} & $898_{-71}^{+67}$    & $6.2_{-0.8}^{+0.9}$\tablenotemark{6}   & $5.4_{-1.0}^{+1.2}$ &$2.70\pm0.08$&  $1.22_{-0.13}^{+0.15}$  & $5.8_{-1.6}^{+2.3}$  \\
XLSS J0223-0436     & 1.22 & $799\pm129$\tablenotemark{e} & $1011_{-79}^{+73}$   & $3.8_{-1.9}^{...}$\tablenotemark{7}  & $2.4_{-1.5}^{...}  $ &$2.46\pm0.06$&  $1.18_{-0.11}^{+0.12}$  & $7.4_{-1.8}^{+2.5}$  \\
RDCS J0849+4452     & 1.26 & $720\pm140$\tablenotemark{k} & $740_{-44}^{+41}$    & $5.2\pm1.3$\tablenotemark{8}  & $3.8_{-1.4}^{+1.5}$& $2.55\pm0.05$&  $0.98_{-0.07}^{+0.77}$  & $4.4_{-0.9}^{+1.1}$  \\
RDCS J0910+5422     & 1.11 & $675\pm190$\tablenotemark{l} & $895_{-51}^{+48}$    & $6.4\pm1.4$\tablenotemark{8}  & $7.4_{-2.3}^{+2.6}$ & $2.61\pm0.05$&  $1.07_{-0.07}^{+0.08}$  & $5.0_{-1.0}^{+1.2}$  \\
RDCS J1252-2927     & 1.24 & $747_{-84}^{+74}$\tablenotemark{m} & $957_{-48}^{+45}$   &$7.6\pm1.2$\tablenotemark{8}        & $4.4_{-1.0}^{+1.1}   $ & $2.47\pm0.03$&  $1.14_{-0.06}^{+0.06}$  & $6.8_{-1.0}^{+1.2}$  \\
XMMU J2235-2557     & 1.39 & $802_{-48}^{+77}$\tablenotemark{n}  & $1145\pm 70$     & $8.6_{-1.2}^{+1.3}$\tablenotemark{9}    & $6.1_{-1.2}^{+1.4}$ & $2.38\pm0.04$&  $1.13_{-0.07}^{+0.08}$  & $7.3_{-1.4}^{+1.7}$  \\
CL J1226+3332       & 0.89 & $1143\pm 162$\tablenotemark{o} & $1237\pm 22$        & $10.4\pm0.6$\tablenotemark{10}           & $12.2_{-1.0}^{+1.1}$& $2.52\pm0.03$&  $1.68_{-0.09}^{+0.10}$  & $13.7_{-2.0}^{+2.4}$ \\
MS 1054-0321        & 0.83 & $1156\pm 82$\tablenotemark{p}  & $1150\pm 50$        & $8.9_{-0.8}^{+1.0}$\tablenotemark{11}    & $12.8_{-1.7}^{+2.2}$& $2.61\pm0.03$&  $1.59_{-0.09}^{+0.11}$  & $10.8_{-1.8}^{+2.1}$ \\ 
CL J0152-1357       & 0.84 & $919\pm 168$\tablenotemark{q}  & $903 \pm 56$        & $6.7\pm1.0$\tablenotemark{8}    & $7.3_{-1.7}^{+1.8}$ & $2.81\pm0.04$&  $1.17_{-0.06}^{+0.09}$  & $4.4_{-0.5}^{+0.7}$ \\
RDCS J0848+4453     & 1.27 & $700\pm 180$\tablenotemark{k}  & $762 \pm 120$       & $3.8\pm1.9$\tablenotemark{8}    & $1.1_{-0.7}^{+0.9}$&$2.57\pm0.05$& $0.96_{-0.07}^{+0.09}$  & $3.1_{-0.8}^{+1.0}$ \\
\enddata

\tablecomments{$^a$ Hilton et al. (2010). 
$^b$ Santos et al. (2009). 
$^c$ Huang et al. (2009). 
$^d$ Eisenhardt et al. (2008).
$^e$ Meyers et al.  (2011).
$^f$ Brodwin et al. (2011).
$^g$ Andreon et al. (2008).
$^h$ Cain et al. (2008).
$^i$ Gilbank et al. in prep.
$^j$ Gilbank et al. (2008).
$^k$ Jee et al. (2006).
$^l$ Mei et al. (2006).
$^m$ Demarco et al. (2007).
$^n$ Rosati et al. (2009).
$^o$ Holden et al. (2009).
$^p$ Tran et al. (2007).
$^q$ Demarco et al. (2005).
$^1$ Hilton et al. (2010).
$^2$ Santos et al. (2009).
$^3$ Maughan et al. (2006).
$^4$ Andreon et al. (2011).
$^5$ Cain et al. (2008).
$^6$ Hicks et al. (2008).
$^7$ Bremer et al. (2006).
$^8$ Ettori et al. (2009).
$^9$ Rosati et al. (2009).
$^{10}$ Maughan et al. (2007).
$^{11}$ Jee et al. (2005b).
}
\end{deluxetable}

\begin{deluxetable}{lcc}
\tabletypesize{\scriptsize}
\tablecaption{Discovery Probability of Galaxy Clusters}
\tablenum{3}
\tablehead{\colhead{Cluster name} &  \colhead{Within Parent Survey} & \colhead{100 deg$^2$}   }
\tablewidth{0pt}
\startdata
XMMXCS J2215-1738	&   	0.96	&	1\\
XMMU J2205-0159        	&	1	&	1\\   
XMMU J1229+0151       	&	0.61	&	0.81\\
WARPS J1415+3612     	&	0.65	&	0.96\\
ISCS J1432+3332          	&	0.14	&	0.77 \\  
ISCS J1429+3437         	&	0.15 &       0.79 \\
ISCS J1434+3427         	&	1	&	1\\
ISCS J1432+3436         	&	0.11	&	0.70 \\
ISCS J1434+3519     	&	1       &	1  \\
ISCS J1438+3414     	&	0.92 	&  	1  \\
RCS 0220-0333       		&	0.74	&	0.91  \\
RCS 0221-0321       		&	1	& 	1    \\
RCS 0337-2844       		&	0.84	&	1    \\                 
RCS 0439-2904       		&	0.95	&	1 \\
RCS 2156-0448       		& 	1	&	1	 \\
RCS 1511+0903       	& 	1	&	1	 \\
RCS 2345-3632       		&	1 	&	1	 \\
RCS 2319+0038       	& 	0.83	&	0.93   \\
XLSS J0223-0436     	&	0.01	&	0.18   \\
RDCS J0849+4452     	& 	0.03 &	0.60  \\
RDCS J0910+5422     	& 	0.06	&	0.65	\\
RDCS J1252-2927     	& 	0.002&	0.052 \\
XMMU J2235-2557     	&	0.013&	0.024 \\
CL J1226+3332       		&	0.006&    	0.01	\\
MS 1054-0321        		& 	0.35	&	-\tablenotemark{a}    \\
CL J0152-1357       		& 	1 	& 	-\tablenotemark{a}    \\
RDCS J0848+4453     	& 	0.08	&	0.88 \\
\enddata
\tablecomments{$^a$ MS 1054-0321 and CL J0152-1357 were discovered in the Einstein Medium Sensitivity Survey (EMSS), which covered a total of
778 deg$^2$ (Gioia et al. 1990).}
\end{deluxetable}

\clearpage

\appendix

\section{A. PSF Modeling} \label{appendix_psf_modeling}

The importance of accurate removal of PSF effects cannot be stressed too much in weak-lensing analyses. Both
dilution and bias of shear signals must be accounted for with rigor in order to advance from mere detection
into measurement. One difficulty in ACS PSF modeling is the small field of view ($3\arcmin\times3\arcmin$), which
does not provide a sufficient number of high S/N stars. Because ACS PSFs vary with time and position, it
is not possible to interpolate/extrapolate the PSF information reliably based a few tens of stars on a target image.
Fortunately, the PSF pattern in ACS seems to be repeatable. That is, one can find
two or more observations remote in time but closely related to each other in the behavior of the PSF pattern (Jee et al. 2007b).
Therefore, it is possible to use a limited number of stars on one image, 
find another image possessing a similar PSF pattern but instead with many more stars covering the entire field,
and utilize this second image to infer the PSF in the first image.
In practice, a slight complexity arises because in general
a target field is visited with many telescope pointings that are different not only in shift and rotation, but
also in PSF pattern. 
Therefore, one needs to determine PSF patterns for individual pointings first, and then
apply shift and rotation to model the PSFs on the final mosaic image.

The above method has been successfully applied to our previous HST weak-lensing analysis of clusters. We 
refer readers to those papers (e.g., Jee et al. 2009) for details. Here, we only summarize the PSF modeling
procedure for the current data sets. We use the Jee et al. (2007b) PSF library, which provides
a polynomial description of two-dimensional PSF variations based on principal components analysis (PCA).
This PCA method allows us to obtain compact basis functions to describe the spatial variation of PSFs (Jee et al. 2007; Jee \& Tyson 2011).

We identified stars on each cluster observation using objects' size and magnitude.
The best PSF template for each cluster observation frame was determined by comparing both size and ellipticity
of these stars. We stored the name of the template, and the alignment information (i.e., shift in pixels and
rotation angle in degrees). After completing this task, we then went through objects in the source catalog that
is created from the final stack, computed their location within each input frame, and interpolated
the coefficients of the principal components at the location. These coefficients were transformed to
two-dimensional PSF images, which were stacked to generate the final PSF model after proper rotation
was applied; in fact, the charge transfer inefficiency (CTI) correction (see \textsection\ref{appendix_cti_correction})
was made prior to stacking by stretching the individual PSFs in the direction of CTI trail.

\begin{figure}
\includegraphics[width=14cm]{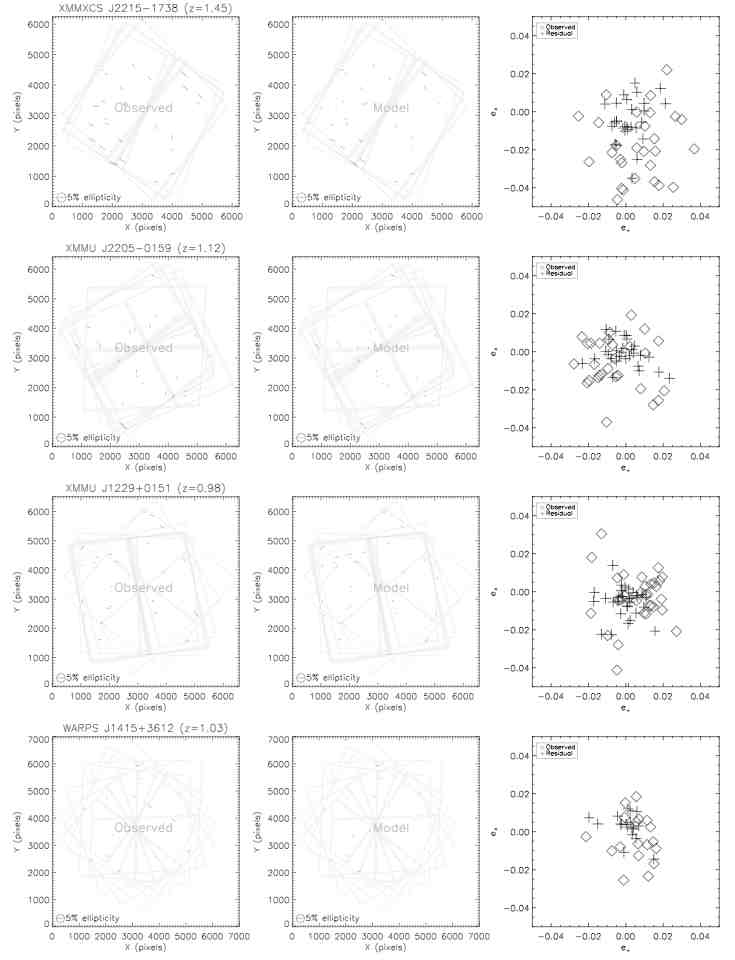}
\caption{Point Spread Function correction. The left panels show the observed ellipticity pattern of the stars whereas in the middle
panels we display the reconstructed ellipticity pattern at the location of the stars. The right panels compares the two ellipticity
components $e_{+}$ and $e_{\times}$ before and after the correction.
\label{fig_psf_page1}}
\end{figure}

\begin{figure}
\includegraphics[width=14cm]{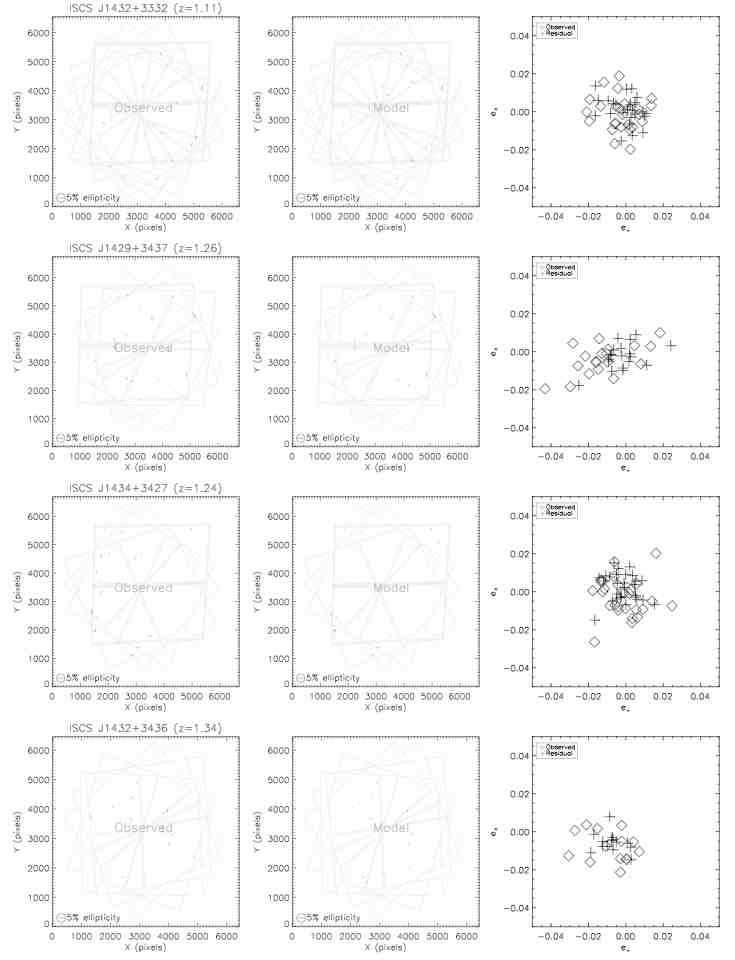}
\caption{Point Spread Function correction - continued from Figure~\ref{fig_psf_page1}
\label{fig_psf_page2}}
\end{figure}

\begin{figure}
\includegraphics[width=14cm]{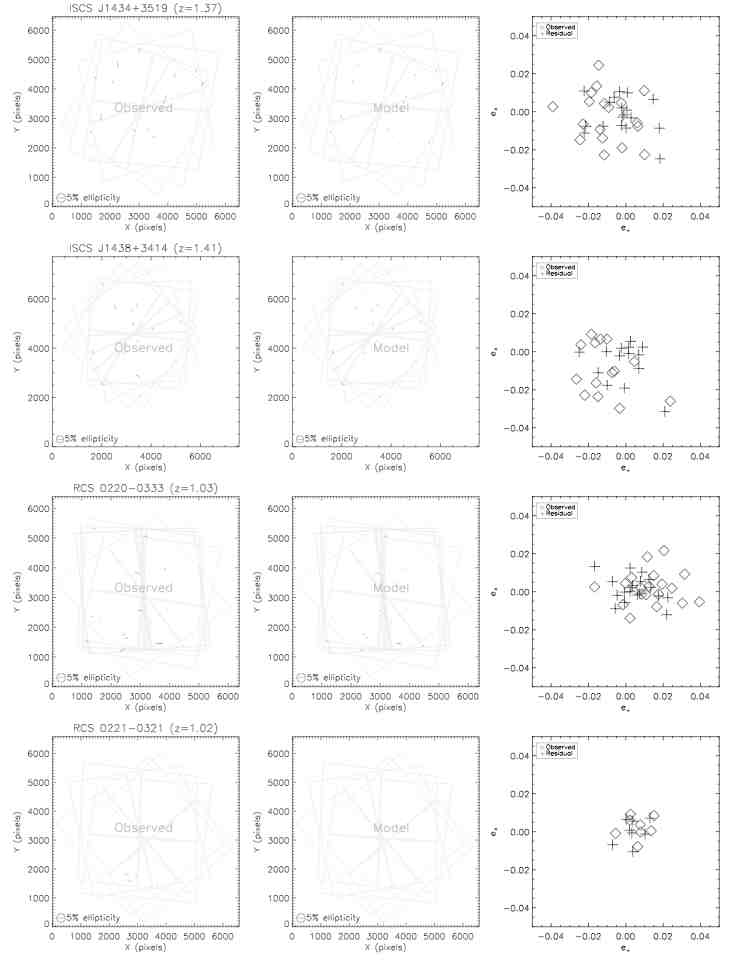}
\caption{Point Spread Function correction - continued from Figure~\ref{fig_psf_page2}
\label{fig_psf_page3}}
\end{figure}

\begin{figure}
\includegraphics[width=14cm]{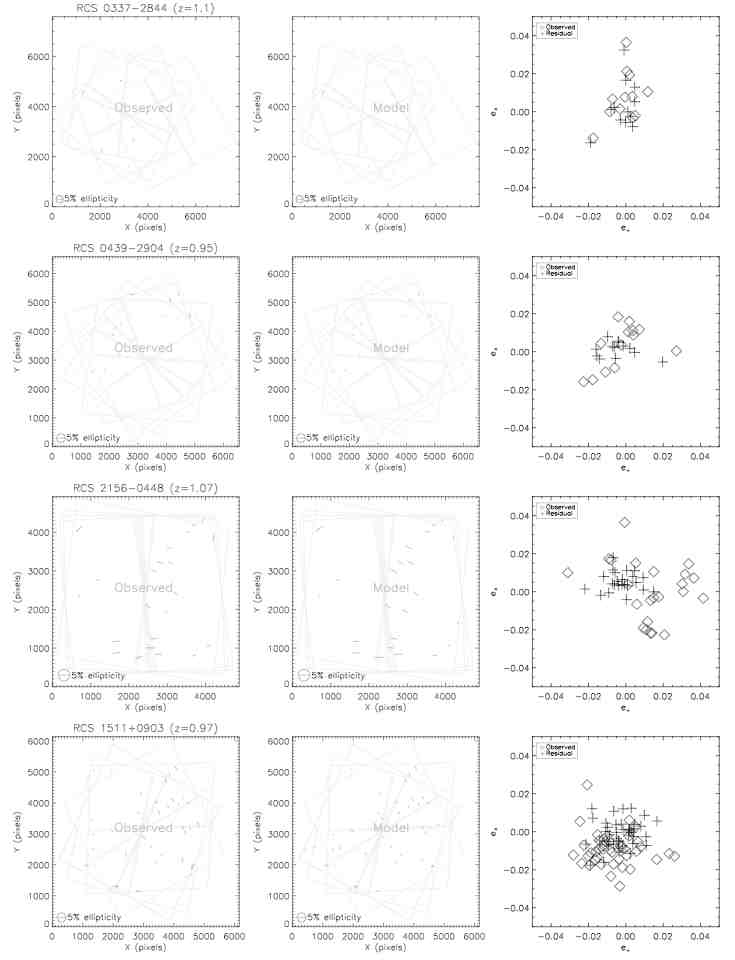}
\caption{Point Spread Function correction - continued from Figure~\ref{fig_psf_page3}
\label{fig_psf_page4}}
\end{figure}

\begin{figure}
\includegraphics[width=14cm]{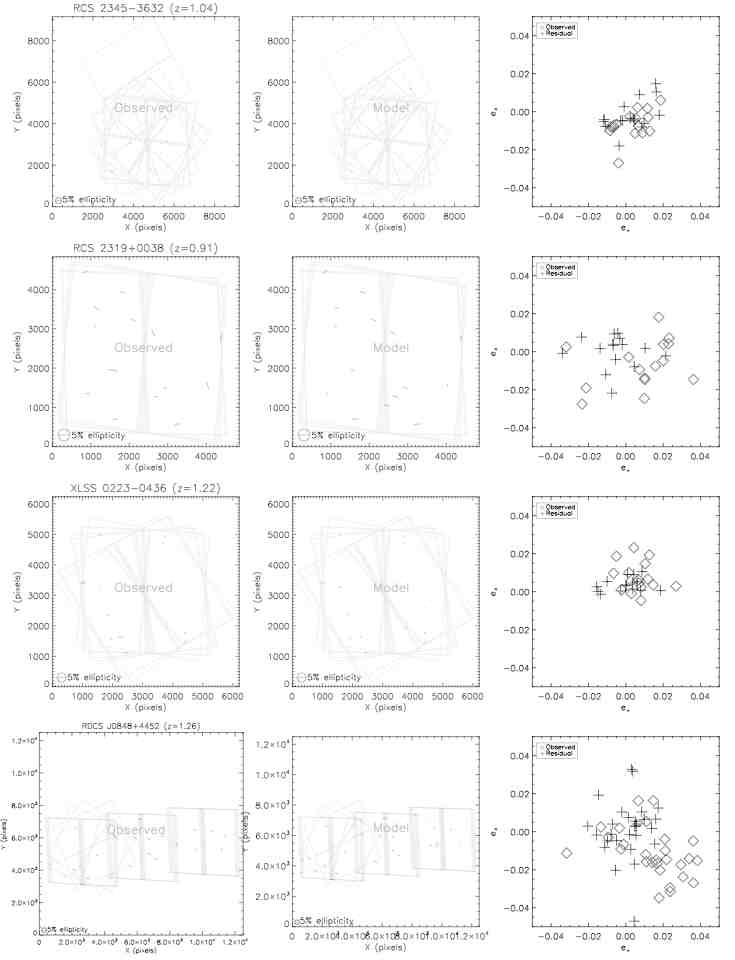}
\caption{Point Spread Function correction - continued from Figure~\ref{fig_psf_page4}
\label{fig_psf_page5}}
\end{figure}

\begin{figure}
\includegraphics[width=14cm]{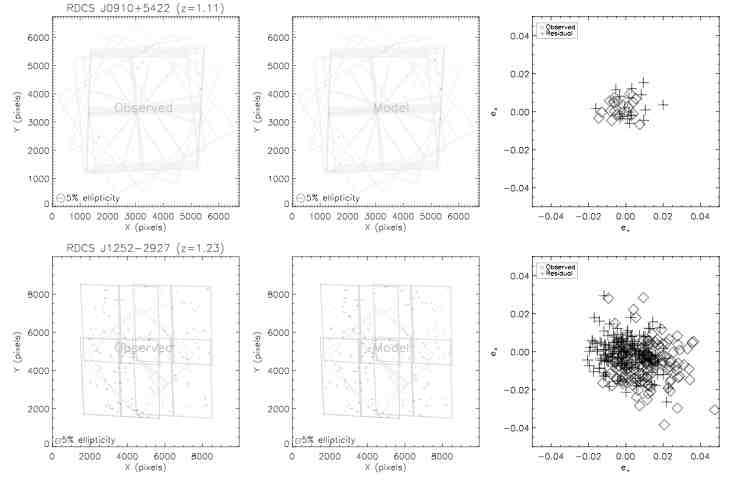}
\caption{Point Spread Function correction - continued from Figure~\ref{fig_psf_page5}
\label{fig_psf_page6}}
\end{figure}

In Figure~\ref{fig_psf_page1}-\ref{fig_psf_page6} we show the PSF correction results. We note that the observed PSF ellipticity
on the stack is on average smaller than the typical value because of the rounding
effect from field rotation. The rms residual per cluster is less than 1\%. Despite some occasional small mismatches, the current
level of the PSF correction exceeds the requirement for cluster lensing analysis.

\section{B. CTI correction }\label{appendix_cti_correction}

During a CCD readout, some fraction of the charge is trapped because
of defects in the silicon. These trapped charges are soon
released after a characteristic time $\tau$. This cycle of trap-release events
continues until the charges in the last row are transferred.
This artifact is sometimes clearly visible as ``charge trails''.

The HST/ACS CTI worsens with time because the number of defects in the CCD is proportional
to the amount of the time exposed to the radiation-rich space environment.
Most of the cluster images in our program are obtained during the year 2006 (four years after the installation of ACS), and
thus we expect the CTI to be much worse than in our previous studies.

Obviously, the effect is undesirable in most applications of the data. The effect is particularly
undesirable in weak-lensing measurements, where one is looking for a subtle distortion of small galaxies.
In this analysis, the CTI-induced
elongation must be carefully measured and accounted for.

In the characterization of the CTI, we categorize the methods in the literature into four schemes.

\begin{enumerate}
\item Difference photometry.
\item Pixel Response measurement.
\item Ellipticity bias of astronomical objects.
\item Ellipticity bias of sub-seeing features (e.g., cosmic ray, hot or warm pixels, etc.).
\item Pixel level correction.
\end{enumerate}

Method 1 utilizes the fact that the CTI-induced elongation leaves some fraction of the charge
outside the aperture. Because the fractional loss depends on the CTI, in this scheme an object
should be observed multiple times in different conditions (e.g., background, time, row, column, etc.). 
Therefore, this method requires large amounts of well-planned, dedicated observations (e.g., Riess \& Mack 2004). 
This scheme is also useful when one's interest is limited to photometric correction of the CTI effect.
Method 2 involves uniformly illuminating
a CCD and measuring deviations from that uniformity
in the few pixels farthest from the readout register.
A series of programs for First Pixel Response (FPR) and Extended Pixel Edge Response (EPER) monitoring have
been carried out (Prop. ID 9649, 10044, 10369 and 10732). A similar, but more
elaborate measurement of this kind is possible if the CCD is exposed to a number of point sources (i.e., single pixel events created by
a controlled X-ray source).
However, this radiation test is only possible on the ground.
Method 3 and 4 rely on the fact that for ACS, the CTI-induced elongation is mostly significant along the parallel charge-transfer
direction (i.e., y-axis in CCD coordinates).
Therefore, taking the average over a number of
measurements should reveal the net CTI effect. When employing Method 3, one has to ensure that
the shapes of the employed objects are free from other systematic effects (e.g., Schrabback et al. 2010; Hoekstra et al. 2011).
For example, poor PSF modeling can induce a residual shape bias that is 
similar to the CTI-induced distortion in both direction and flux-dependence.
Method 4 differs from Method 3 in that the ellipticity bias is measured from cosmic rays or
warm pixels, whose shapes are not subject to geometric distortion or PSF. Because these
sub-seeing features are numerous even on a single 500 s exposure image, it is possible to
obtain high S/N information on CTI from the data themselves. We applied this method
to the weak-lensing analysis of the then highest redshift cluster XMM2235 at $z=1.4$ 
(Jee et al. 2009). As Jee et al. (2009) were somewhat implicit on how to relate the measured ellipticity 
bias of cosmic rays to object ellipticity, we provide details in the current paper.
Method 5 is recently suggested by Massey et al. (2010) and Anderson \& Bedin(2010). This approach aims to restore
CTI-trails back to the original pixels based on profile measurements of warm-pixels. Potentially, this method provides 
convenient CTI-corrected images, which astronomers can work on directly without worrying about the details of the model. However,
The fidelity of the method has yet to be tested.

\subsection{B.1 Quantification of CTI Effect}

Any significant sub-seeing features (hereafter SSFs) present in the raw CCD images are mostly 
cosmic rays, warm/hot pixels, or simply photon noise,
which are commonly removed through elaborate image-processing.
However, knowing that these SSFs are not caused by the photons following the regular
optical path of the instrument, we can use these otherwise nuisance features to our advantage.
Because position angles of the SSFs in principle should possess no preferred direction on the
surface of the CCD, any deviation from the anticipated isotropy is a clean indicator of
the artifact in CCD readout.

We used {\tt SExtractor} to detect SSFs by
looking for objects whose half-light radius is less than 1 pixel. We did not let SExtractor filter
the images with matched-PSFs, and thus most of these SSFs would have occupied $\sim1$ pixel
if there had been no CTI-trails. We also measured the ellipticity of the SSFs by
evaluating the unweighted moments using the unfiltered images. This ellipticity
measurement scheme is different from Jee et al. (2009), where we
measured the ellipticity after the image was convolved with a Gaussian function
matching the instrument PSF. The qualitative behavior of the CTI 
does not change by this difference in measurement scheme. However, the
current measurement more clearly reveals the "turnaround" feature that we discuss below.

\begin{figure}
\plottwo{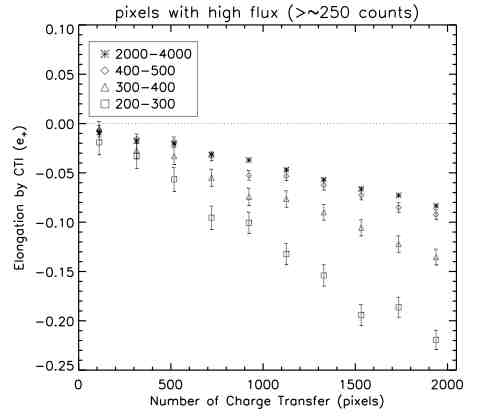}{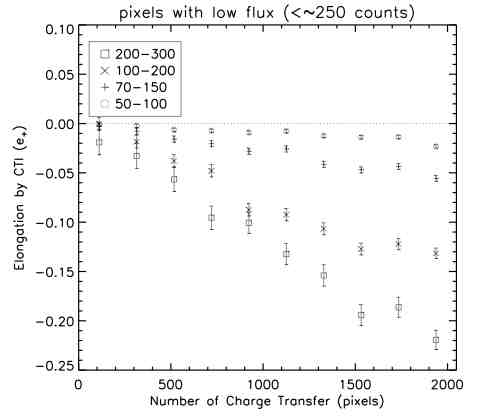}
\caption{CTI trail of SSFs in the XMMXCS J2215-1738 ACS images. The ellipticity bias (i.e., elongation along the parallel readout direction) due to 
CTI linearly increases (becomes more negative) 
with number of charge transfers. We divide the cases into two groups according to their fluxes in order to demonstrate the different flux-dependence.
For the pixels with high counts (left), we observe that the CTI worsens for decreasing fluxes, as was reported in previous studies. However,
the trend is reversed for the pixels with low counts (right).
\label{fig_e_vs_transfer}}
\end{figure}

As a case study, we display the ellipticity bias of the SSFs in 
the galaxy cluster XMMXCS J2215-1738 in
Figure~\ref{fig_e_vs_transfer}. 
The left and right panels show the results for the SSFs,
whose fluxes (counts) are high and low, respectively. Although here we use the SSFs
detected only in the images of XMMXCS J2215-1738, they are still numerous and thus enable us to characterize
the behavior of the CTI with high significance. First, it is clear that the average ellipticity
of the SSFs in each flux range linearly increases (thus more negative) with the number of
charge transfer. Second, for the bright SSFs, the CTI effect worsens for decreasing fluxes.
In contrast, this trend is reversed for low-flux SSFs. This CTI turnaround was first reported
in Jee et al. (2009) through both the SSF analysis and stellar photometry. Schrabback
et al. (2010) also observed this turnaround in their independent CTI analysis. It is worth
noting that the XMMXCS J2215-1738 cluster was observed in the year 2006 in short ($\sim500$ s) exposures, and
the background level is very low ($\sim0$ counts), which explains the severe CTI-trails. 

\begin{figure}
\includegraphics[width=8cm]{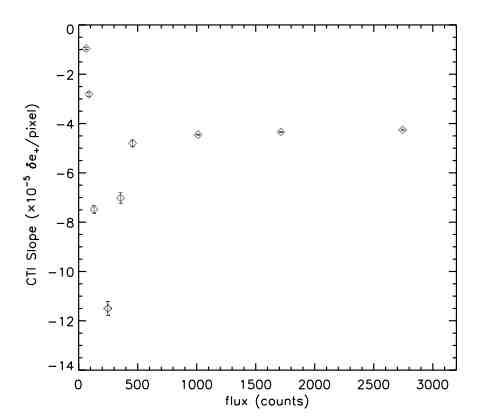}
\caption{CTI slope versus flux. Here we compress the information in the two panels in Figure~\ref{fig_e_vs_transfer} into one by plotting the
slopes as a function of the fluxes. It appears that the flux-dependence can be classified into three regions. In the high-flux regime ($\gtrsim500$),
the slope changes very slowly with flux. In the intermediate-flux regime ($250\lesssim $counts$\lesssim 500$), there shows a precipitous decline of
the CTI slope (more elongation) for decreasing flux. Finally, in the low-flux regime ($\lesssim 250$), a sharp turnaround is observed.
We found that most of our background galaxies belong to the low end of this regime. Hence, the CTI effect is negligible for our source population.
\label{fig_cti_slope}}
\end{figure}

One way to compress the information in the two
panels in Figure~\ref{fig_e_vs_transfer} is to plot the slopes of the data points as a function of
flux as shown in Figure~\ref{fig_cti_slope}\footnote{When the slope is multiplied by 2048, the
result gives the maximum ellipticity bias farthest from the readout register}. 
The plot helps us to
realize how sensitive the CTI slope is to the flux. Especially, it is remarkable that the 
CTI turnaround is very sharp (compare this with Figure 4 of Jee et al. [2009]).

Before we proceed further, it is instructive to estimate what magnitude range of galaxies
is mostly affected by this CTI pattern. As is mentioned in \textsection\ref{section_selection},
we select 24-28 mag galaxies as source objects. Considering that the average exposure
time is $\sim$500 s and the mean FWHM of the source galaxies is significantly
larger than the SSFs ($\sim8$ pixels versus $\sim1.7$ pixels),
we estimate that this magnitude range corresponds to 2-100 counts in Figure~\ref{fig_cti_slope}.
Therefore, the brightest and farthest (from the readout register) galaxies in our source sample suffer 
the CTI degradation most.

\subsection{B.2 CTI Correction to Galaxies}

We characterize the CTI with the ellipticity bias of the SSFs. Then, a nontrivial question is
how to quantitatively relate this ellipticity bias to galaxy ellipticity change. Obviously, the
ellipticity bias in SSF is a significant exaggeration of what occurs to galaxy shapes;  a 
small CTI-trail induces much larger ellipticity to a $\delta$ function-like feature
than to, for example, a $0.5\arcsec$ galaxy. The exact translation requires
our knowledge on the shape of the CTI-induced trail and the surface brightness profile
of galaxies. Consequently, the recovery of the galaxy shape prior to
the CTI effect involves the same delicacies in PSF-effect correction, which is a more familiar
problem.

Indeed, the CTI trailing can be treated as convolution with a one-dimensional kernel,
which further smears the post-seeing objects, but only along the readout direction.
\begin{equation}
o(x,y)=i(x,y)\otimes p(x,y) \otimes c(x,y) = i(x,y) \otimes r(x,y)
\end{equation}
\noindent
where $i(x,y)$ and $o(x,y)$ are the intrinsic and observed images, respectively. 
$p(x,y)$ and $c(x,y)$ are the convolution kernel representing the PSF and the CTI-trailing, respectively.
Because the associative law holds for convolution, the smearing of
the object can be viewed as a single convolution by the kernel $r(x,y)$, which
includes both the PSF and CTI-trailing. One caveat here is that the CTI kernel
$c(x,y)$ in practice depends on flux, and, because a galaxy image consists
of multiple pixels of varying flux, it is essential to approximate the effect with a single kernel, representing
the collective effect of CTI on each galaxy.

One can determine the shape of the kernel $c(x,y)$ by stacking many CTI-trails. We find that
on average at short distances, CTI trails are well-approximated by an exponential tail $\propto e^{-y/\tau}$, where
the time constant $\tau$ (expressed in units of pixel) determines how quickly the trail drops.
Upon close examination, however, we observe that the measured CTI-trails differ in that
1) at large distances ($>$ a few pixels) the measured CTI-trail slope is shallower, and
2) occasionally a couple of small bumps appear, which suggests that different species of
charge traps might be present.
Nevertheless, we chose the simple functional form of a single exponential because object shapes are
most sensitive to the information in the central few pixels.
Based on the laboratory experiments of Dawson et al. (2008), Rhodes et al. (2010) also concluded that object shapes are most sensitive to the first few pixels and 
a single exponential term is adequate to describe the trails.

\begin{figure}
\includegraphics[width=8cm]{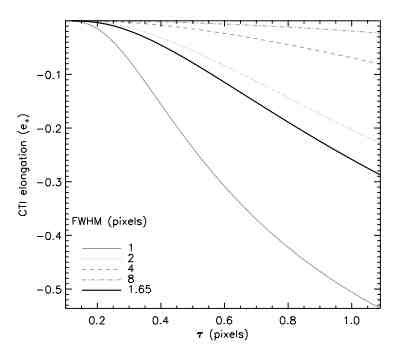}
\caption{Time constant versus elongation. We model the CTI trail with $\propto e^{-\tau}$, where the time constant $\tau$ determines how
fast the trail truncates. The resulting ellipticity also depends on the size of the SSFs; for example, more ellipticity is induced for cosmic rays that occupy
less number of pixels on average. The difference lines represent the results for different sizes. The thick solid line shows the case
for the FWHM value of 1.65, which matches the mean size of the SSFs that we use for the study.
\label{fig_ssf_trail_simulation}}
\end{figure}

\begin{figure}
\includegraphics[width=8cm]{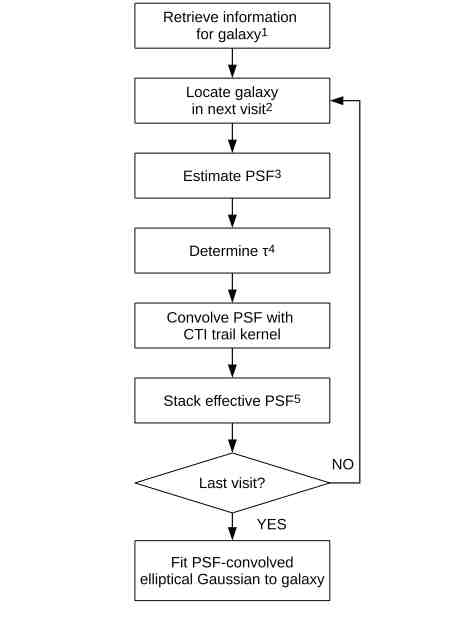}
\caption{Galaxy shape measurement procedure with CTI and PSF corrections. 1. Surface brightness and coordinate
are retrieved from the master source catalog. 2. Coordinate transformation is performed to locate
the object in each visit's detector coordinate system. 3. Using the PSF model for each visit, we
estimate the PSF at the object's location. 4. Based on the y-coordinate (i.e., number of charge transfer pixels)
and the surface brightness, we determine the ellipticity bias (e.g., using the relation in Figure~\ref{fig_cti_slope}). Then, this
ellipticity bias $\delta e_{+}$ is converted to the CTI time constant $\tau$ (e.g., using the relation in Figure~\ref{fig_ssf_trail_simulation}).
5. The effective PSF (modified with CTI information) is rotated and scaled according to the exposure time prior
to stacking for the creation of the final PSF.
\label{fig_flowchart_cti}}
\end{figure}

Now the remaining question is how to determine the relation between the ellipticity bias
$\delta e_{+}$ in the SSFs and the CTI time constant $\tau$. For this investigation, we
rely on image simulations, where we create many artificial SSFs features and trail
them with the $\sim e^{-y/\tau}$ kernel. Because we select only the $r_h\sim1$ 
SSFs to perform  the study above,
no exhaustive efforts are required to match the artificial SSFs to the real ones.
Figure~\ref{fig_ssf_trail_simulation} shows the $\tau$ vs. SSF $ellipticity$ bias
calibration that we derived from this image simulation. The thick solid line represents
the relation when the size of the simulated SSFs matches the observed ones. The other
lines illustrate how much the slope of the CTI trail depends on object
sizes.
The flowchart in Figure~\ref{fig_flowchart_cti} summarizes the procedure to
correct the PSF and CTI effects.

\begin{figure}
\includegraphics[width=8cm]{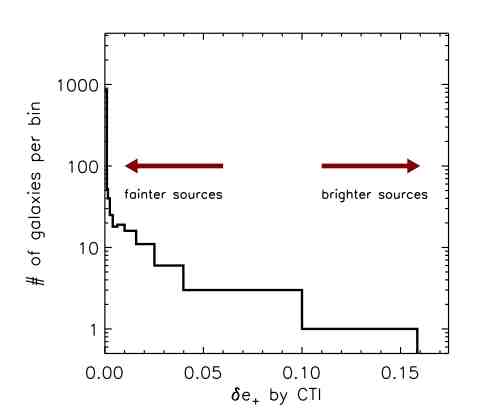}
\caption{Number of galaxies affected by CTI elongation. We examine how many background galaxies in the ACS field are affected by 
a given amount of CTI elongation $\delta e_{+}$. Very few galaxies are subject to large elongation ($\gtrsim 0.05$).
A majority of galaxies, which fall to the CTI-mitigation regime are elongated by $\delta e_{+}<<0.01$, which can be safely ignored for cluster lensing analysis.
\label{fig_e_change_by_cti}}
\end{figure}

Although the CTI correction described here is complicated and the result of
time-consuming effort, we emphasize that the effect is on average small
(also perhaps negligible in many cases) for cluster lensing analysis.
Figure~\ref{fig_e_change_by_cti} shows that very few galaxies are subject to large elongation ($\gtrsim 0.05$).
A majority of galaxies, which fall to the CTI-mitigation regime are elongated by $\delta e_{+}<<0.01$.

\end{document}